%% file: main.tex
\documentclass[11pt]{article}
\usepackage{jheppub}
\usepackage{amsmath,amssymb,amsfonts,graphicx,slashed,color, mathtools, amsthm}
\usepackage{bbold}
\usepackage{comment}
\usepackage{enumerate}
\usepackage{float}
\usepackage{caption}
\usepackage{subcaption}
\usepackage{pdflscape}
\usepackage{physics}
\usepackage{tikz}
\usepackage{mathrsfs}
\usepackage[dvipsnames]{xcolor}
\usepackage{pgfplots}
\usetikzlibrary{arrows.meta, decorations.markings}
\tikzset{
    pil/.style={->, thick, shorten <=2pt, shorten >=2pt,},
    mid arrow/.style={postaction={decorate,decoration={
        markings,
        mark=at position .5 with {\arrow[#1]{stealth'}}
    }}},
}
\pgfplotsset{compat=1.9}

\newcommand{\hs}{\mathcal{H}} 

\newcommand\numberthis{\addtocounter{equation}{1}\tag{\theequation}}

\definecolor{Kbrown}{HTML}{8B4513}
\definecolor{Kblue}{HTML}{007Aff}
\definecolor{Kgreen}{HTML}{94E700}
\definecolor{Korange}{HTML}{FF8800}
\definecolor{Kred}{HTML}{FF0000}
\definecolor{Kpink}{HTML}{FF9797}
\definecolor{Kdarkblue}{HTML}{03468F}
\definecolor{Kdarkgreen}{HTML}{007355}
\definecolor{Kpurple}{HTML}{82218b}

\title{Subregion observer rules from generalized entanglement wedges}

\author[a,b]{Lana Bozanic,}
\author[a]{Kenneth Higginbotham,}
\author[a]{Chris Waddell}

\affiliation[a]{Perimeter Institute for Theoretical Physics, 31 Caroline St N, Waterloo, ON\ N2L 2Y5, Canada}
\affiliation[b]{Department of Physics and Astronomy, University of Waterloo, 200 University Ave W, Waterloo, ON, N2L 3G1, Canada}
\emailAdd{lbozanic@perimeterinstitute.ca}
\emailAdd{khigginbotham1@perimeterinstitute.ca}
\emailAdd{cwaddell@perimeterinstitute.ca}

\abstract{
We consider rules for modifying holographic tensor networks proposed in two independent contexts: by the Colorado (CO) group \cite{akers_observers_2025} to incorporate observers in holographic maps, and by Kaya-Rath-Ritchie (KRR) \cite{kaya_hollowgrams_2025} to derive the Bousso-Penington generalized entanglement wedge proposal. Interestingly, these two sets of tensor network rules are exactly equivalent. This suggests a more general connection between these Abdalla-Antonini-Iliesiu-Levine (AAIL) inspired observer rules and generalized entanglement wedges. To pursue this connection, we first use KRR's analogous rules for the gravitational path integral (based on fixed geometry states) to generalize AAIL's path integral rules to include observers occupying a bulk subregion. Additionally, we leverage the connection in the opposite direction by using the AAIL rules to derive the Bousso-Penington proposal for pointlike bulk regions in JT gravity.
}

\begin{document}

\maketitle

\section{Introduction}

In lieu of experimental evidence for a theory of quantum gravity, great progress has been made by investigating paradoxes within our current understanding of gravity and quantum mechanics. Famous among these is of course the black hole information paradox \cite{hawking_particle_1975}, which led to new innovations in tools such as the gravitational path integral \cite{almheiri_replica_2020,penington_replica_2020} and non-isometric codes \cite{akers_black_2024}. These same tools have led us to new paradoxes in closed universes in the form of a trivial fundamental Hilbert space ($\dim\hs_\text{CU} = 1$)
\cite{maldacena_wormholes_2004,
almheiri_page_2020,
penington_replica_2020,
marolf_transcending_2020,
mcnamara_baby_2020,
usatyuk_closed_2024,
usatyuk_closed_2025} 
and the Antonini-Rath puzzle \cite{antonini_holographic_2025}, providing fertile ground for new understandings of holography and quantum gravity.

One possible solution (among many) to these closed universe puzzles is to make new modifications to our path integral and non-isometric code tools. Two proposals have been motivated by the incorporation of bulk observers:
\begin{itemize}
    \item Harlow-Usatyuk-Zhao (HUZ) \cite{harlow_quantum_2026}, motivated by the assumption that observers are classical;
    \item Abdalla-Antonini-Iliesiu-Levine (AAIL) \cite{abdalla_gravitational_2025}, motivated by the assumption that relational observables require the observer to be present in their own universe
\end{itemize}
Both works proposed modifications to the path integral based on their definition of an observer. By cloning the observer to an external reference, HUZ suppressed certain terms in the path integral by the dimension of the observer's Hilbert space ($d_\text{ob} \equiv \dim\hs_\text{ob}$). Alternatively, AAIL required that an observer's worldline must connect their corresponding bra and ket, thereby entirely removing these same terms from the path integral. In the context of JT gravity, both found non-trivial dimensions for a closed universe Hilbert space:
\begin{equation} \label{eq:dimH_HUZ_AAIL}
    \dim\hs_\text{CU}^\text{(HUZ)} \approx \min (d_\text{ob},\,e^{2S_0}) \: , \qquad \dim\hs_\text{CU}^\text{(AAIL)} \approx d_\text{ob} e^{2S_0} \: .
\end{equation}

In addition to their path integral modifications, HUZ also provided corresponding rules for incorporating observers into non-isometric codes, by again cloning the observer to an external reference. To complete the comparison, the Colorado (CO) group proposed a set of non-isometric code rules analogous to AAIL's path integral rules \cite{akers_observers_2025}. Motivated by the assumption that the observer is already a part of the fundamental description, the CO rules removed any part of the non-isometric code acting locally on the observer; in a tensor network model for a closed universe, these rules remove any tensors acting on the observer, leaving dangling legs across the observer's boundary. These dangling legs permit an interpretation of the $e^{2S_0}$ contribution to $\dim\hs_\text{CU}^\text{(AAIL)}$ as an area term:
\begin{equation} \label{eq:dimH_CO}
    \dim\hs_\text{CU}^\text{(CO)} \approx d_\text{ob} e^{\mathcal{A}_\text{ob}/4G_N}.
\end{equation}

Independent from the above work on closed universes, entanglement wedges defined from the quantum extremal surface prescription in AdS/CFT
\cite{ryu_holographic_2006,
hubeny_covariant_2007,
faulkner_quantum_2013,
engelhardt_quantum_2015}
also proved essential to our understanding of holography and the black hole information paradox \cite{penington_entanglement_2020,almheiri_entropy_2019}. In an attempt to generalize these results beyond AdS/CFT, Bousso and Penington (BP) proposed a generalized entanglement wedge prescription for assigning an entanglement wedge to any gravitational subregion \cite{bousso_entanglement_2023,bousso_holograms_2023}. More recently, work by Kaya, Rath, and Ritchie (KRR) sought to derive the BP proposal from both random tensor networks and the gravitational path integral \cite{kaya_hollowgrams_2025}. In tensor networks, KRR defined a ``hollowing'' procedure that removes tensors from a bulk subregion, leaving dangling legs whose entropy defines a generalized entanglement wedge.\footnote{A similar tensor network model was used by Sahu, van der Heijden, Van Raamsdonk, and Zibakhsh as a way to understand the algebras associated to generalized entanglement wedges \cite{sahu_algebras_2025}.} Using fixed geometry states \cite{Penington:2022dhr} (a generalization of the fixed area states defined in \cite{akers_holographic_2019,dong_flat_2019}), KRR proposed analogous hollowing modifications to the path integral rules that can be used to identify the generalized entanglement wedge of a specified subregion.

One might suspect that these two parallel developments are related; indeed, observers are naturally associated with an algebra of accessible bulk observables 
\cite{chandrasekaran_algebra_2023,
witten_algebras_2023,
witten_backgroundindependent_2024,
vuyst_gravitational_2025,
kudler-flam_algebraic_2024,
chen_clock_2024,
kolchmeyer_chaos_2024,
maldacena_real_2024,
tietto_microscopic_2025}, and a connection between the CO observer rules and generalized entanglement wedges was hinted at in \cite{akers_observers_2025}. This observation is strengthened by noting that the CO observer rules and KRR tensor network rules \textit{are exactly equivalent}: both remove tensors in some subregion of the network. That is, KRR = CO.\footnote{We do not claim originality of this observation; the equivalence was already noted by KRR in \cite{kaya_hollowgrams_2025}.} Consequently, it is natural to expect a similar relationship between the path integral rules proposed by AAIL for observers and those of KRR for generalized entanglement wedges. In this work, we extend this comparison in two ways: first by using the KRR rules to generalize the AAIL observer rules to include observers associated with a bulk subregion (rather than a worldline), and second by using the AAIL rules to derive the BP proposal in the case of a pointlike region.

There is still a great deal to understand about these observer rules, and we recognize their speculative nature. A priori, we do not have a first principles motivation for preferring the AAIL rules or the HUZ rules; it may even be that neither are sensible, or that both are sensible but compute different quantities. However, observing that the AAIL rules are equivalent to derivations of the BP proposal might help us to interpret what these rules are actually letting us calculate. For example, Balasubramanian and Cummings obtained the BP proposal by carefully accounting for gravitational edge modes in the path integral \cite{balasubramanian_entropy_2024}, highlighting the principal role being played by the gravitational dressing of the subregion. This web of relationships may help to clarify the interpretation of the AAIL Hilbert space.

The remainder of this work will be organized as follows. We begin with a review of the CO and KRR tensor network rules in section \ref{sec:KRR=CO} to demonstrate their equivalence -- readers familiar with both sets of tensor network rules may safely skip this section. In section \ref{sec:GPI}, we will apply KRR's path integral rules to an observer in a closed universe by calculating the generalized entropy of the subregion they occupy. As expected, we find zero when the bulk state is pure, but we can find a non-zero answer by including bulk matter in a sufficiently high entropy mixed state. For concreteness, we demonstrate this in JT gravity as a toy example in section \ref{sec:Sgen_obs_JT}. All calculations in section \ref{sec:GPI} are meant to parallel those done by KRR in sections 4 and 5 of \cite{kaya_hollowgrams_2025} for a subregion of a universe with boundary.

In section \ref{sec:bulk_HS}, we estimate the dimension of the fundamental Hilbert space of the closed universe in the presence of an observer using KRR's rules. We will estimate the dimension in two ways. The first uses the entropy of an external reference entangled with the bulk effective degrees of freedom, similar to the approach used in appendix B of \cite{akers_observers_2025}. The second uses the variation of the inner product,
\begin{equation} \label{eq:variation}
    \sigma^2 = \overline{|\langle\phi|\psi\rangle|^2} - \Big|\overline{\langle\phi|\psi\rangle}\Big|^2,
\end{equation}
as used for example in \cite{harlow_quantum_2026,abdalla_gravitational_2025,akers_observers_2025}. In the case of JT gravity, we find agreement with AAIL's result (\ref{eq:dimH_HUZ_AAIL}). While this result is not new, we find that this application of KRR's rules makes it clear that the $e^{2S_0}$ contribution in (\ref{eq:dimH_HUZ_AAIL}) can be viewed as an area term arising from a conical defect in the path integral, as expected from the CO result (\ref{eq:dimH_CO}).

Section \ref{sec:BP_point} reverses the comparison by using the AAIL path integral rules to find the generalized entanglement wedge of a pointlike subregion in JT gravity. Finally, we conclude with a few remarks in section \ref{sec:conc}.

\subsection{Relation to other work}

Others have considered special treatments of observable subregions within the gravitational path integral. For example, Ivo-Li-Maldacena defined a ``no boundary density matrix'' by tracing out boundary conditions outside of an observable subregion \cite{ivo_no_2024}. A similar ``partial observability'' proposal by Nomura-Ugajin traces out degrees of freedom inaccessible from a particular subregion \cite{nomura_nonperturbative_2025,nomura_physical_2026}. The work \cite{Dong:2020uxp} also provided a prescription for associating an entropy with gravitating subregions, including in the case of closed universes, though this was interpreted as counting low-energy degrees of freedom in the subregion and differs from the Bousso-Penington proposal. While the work by Balasubramanian and Cummings \cite{balasubramanian_entropy_2024} mentioned above does not rely on observability to define a subregion, their techniques could also be applied in this case.

These approaches differ from KRR's treatment of subregions in the path integral, which relies on fixed geometry states. We find here that the analogy between fixed geometry states and tensor networks is helpful for sharpening the connection between the AAIL and CO observer rules. It would be interesting to better understand the connection between KRR's fixed geometry prescription and the above work.

We also note that work by Chen \cite{chen_observers_2025} used ``partial sources'' as a unifying framework for both the HUZ and AAIL modifications to the path integral. It would be interesting to understand if KRR's hollowing procedure could also be included within this framework. 

\textit{Note added:} In preparation of this manuscript, two papers \cite{bousso_quantum_2026,wei_pure_2026} were posted to the arXiv, both using path integral techniques (distinct from the fixed geometry states used here) to study the quantum states associated with generic bulk subregions. While they do not make a connection with the above observer rules, they have potentially interesting overlap with our current work.

\section{KRR = CO} \label{sec:KRR=CO}

In this section, we will spell out the connection between the the CO tensor network observer rules articulated in \cite{akers_observers_2025} and the derivation by KRR \cite{kaya_hollowgrams_2025} of the generalized entanglement wedge proposal of Bousso and Penington \cite{bousso_entanglement_2023, bousso_holograms_2023}. At a high level, we can think of the calculations within all of these frameworks as promoting some subset of the bulk effective degrees of freedom to fundamental degrees of freedom. 

Tensor networks have long been understood to provide a good model for holographic maps \cite{Swingle:2009bg, Pastawski:2015qua, Hayden:2016cfa}, with analogues of the Ryu-Takayanagi formula and entanglement wedge reconstruction. Typically, we think of them as maps from some code subspace of the full bulk Hilbert space, corresponding to the bulk effective Hilbert space on a fixed background, into some fundamental Hilbert space, i.e.
\begin{equation}
    V_{\text{TN}} : \mathcal{H}_{\text{eff}} \rightarrow \mathcal{H}_{\text{fun}} \: .
\end{equation}

The map $V_{\text{TN}}$ has an internal structure which reflects the locality of the bulk effective field theory. To construct it, we may consider some graph $\mathcal{G} = (\mathcal{V}, \mathcal{E})$ representing a discretized version of the geometry on a given time slice $\Sigma$. 
The edges $e \in \mathcal{E}$ of the graph emanating from vertex $x$ are associated with Hilbert spaces $\mathcal{H}_{x, e}$ of dimension $d_{e}$. The vertices of the graph will be associated with tensors $\{T_{x}\}_{x \in \mathcal{V}}$. If there are $k$ edges $e_{1}, \ldots, e_{k}$ incident on a given vertex $x \in \mathcal{V}$, then we demand that $T_{x}$ has index structure
\begin{equation}
    T_{x} = (T_{x})_{\mu_{1} \ldots \mu_{k} b B} \: ,
\end{equation}
and define the (possibly unnormalized) state
\begin{equation}
    | T_{x} \rangle \equiv (T_{x})_{\mu_{1} \ldots \mu_{k} b B} | \mu_{1} \rangle_{x, e_{1}} | \mu_{2} \rangle_{x, e_{2}} \ldots | \mu_{k} \rangle_{x, e_{k}} | b \rangle_{x, \text{bulk}} | B \rangle_{x, \partial} \: ,
\end{equation}
with $d_{e_{i}}$ the dimension of $\mathcal{H}_{x, e_{i}}$, $d_{x, \text{bulk}}$ the dimension of $\mathcal{H}_{x, \text{bulk}}$, and $d_{x, \partial}$ the dimension of $\mathcal{H}_{x, \partial}$. We will interpret the Hilbert space $\mathcal{H}_{x, \text{bulk}}$ as a factor of the bulk effective theory associated with degrees of freedom at $x$, whereas we will interpret the Hilbert space $\mathcal{H}_{x, \partial}$ as a factor of the boundary Hilbert space (this is allowed to be one-dimensional for the tensors in the interior of the network). 

Different choices for $T_x$ correspond to different codes; here we choose each tensor to be random \cite{Hayden:2016cfa},
\begin{equation}
    |T_x\rangle = U_x |0\rangle
\end{equation}
where $U_x$ is a Haar random unitary. In the case of closed universes below, we will replace $U_x$ with a Haar random orthogonal matrix $O_x$,
\begin{equation}
    |T_x\rangle = O_x |0\rangle, \qquad \text{(closed universe)}
\end{equation}
following \cite{harlow_gauging_2023,harlow_quantum_2026,akers_observers_2025}.

\begin{figure}
\centering

\begin{subfigure}{0.5\textwidth}
  \includegraphics[width = \textwidth]{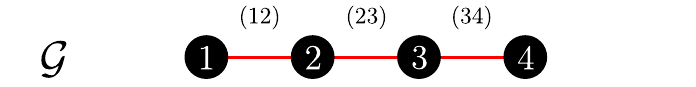}
  \caption{}
  \label{subfig:generic_TN_1}
\end{subfigure}\hfill
\begin{subfigure}{0.5\textwidth}
  \includegraphics[width = \textwidth]{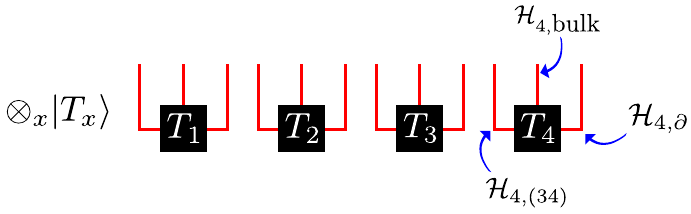}
  \caption{}
  \label{subfig:generic_TN_2}
\end{subfigure}\hfill

\bigskip

\begin{subfigure}{0.5\textwidth}
  \includegraphics[width = \textwidth]{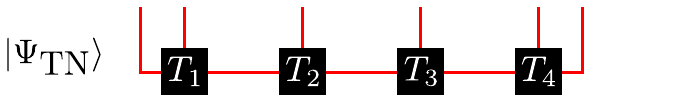}
  \caption{}
  \label{subfig:generic_TN_3}
\end{subfigure}\hfill
\begin{subfigure}{0.5\textwidth}
  \includegraphics[width = \textwidth]{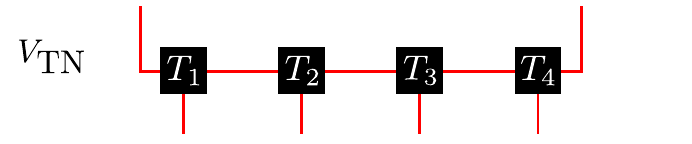}
  \caption{}
  \label{subfig:generic_TN_4}
\end{subfigure}\hfill

\caption{(a) A four-node graph $\mathcal{G}$ representing a spatial slice with the topology of a one-dimensional line. (b) The uncontracted tensor network state $\otimes_{x} | T_{x} \rangle$. The edge, bulk, and boundary Hilbert space factors associated with node $x=4$ are indicated. (c) The contracted tensor network state $| \Psi_{\text{TN}} \rangle$. (d) When the bulk legs are transformed from output legs to input legs by post-selecting on copies of $\langle \text{MAX} |$, we obtain the bulk-to-boundary map $V_{\text{TN}}$. }
\label{fig:generic_TN}

\end{figure}

Having consistently associated tensors with vertices in the graph, we can then construct the state of the ``contracted'' tensor network
\begin{equation} \label{eq:contracted}
    | \Psi_{\text{TN}} \rangle \equiv \left( \otimes_{(xy) \in \mathcal{E}} \langle \text{MAX} |_{xy} \right) \left( \otimes_{x \in \mathcal{V}} | T_{x} \rangle \right) \: ,
\end{equation}
where $| \text{MAX} \rangle_{xy} \equiv d_e^{-1/2}\sum_{\mu} | \mu \rangle_{x, (xy)} | \mu \rangle_{y, (xy)}$ denotes the maximally entangled state. $|\Psi_\text{TN}\rangle$ is a state of both the bulk effective Hilbert space and the boundary Hilbert space, and post-selecting the bulk legs $b_x$ on a bulk state gives a state of the boundary,
\begin{equation}
    |\Psi\rangle_\partial = \langle \psi_\text{bulk} |\Psi_\text{TN}\rangle
\end{equation}
Alternatively, $|\Psi_\text{TN}\rangle$ can be made into a bulk-to-boundary map by projecting the bulk effective degrees of freedom onto states $| \text{MAX} \rangle_{x} \equiv d_{x,\text{bulk}}^{-1/2} \sum_{a} | a \rangle_{x, \text{bulk}} | a \rangle_{x, \text{bulk}'}$ where $\mathcal{H}_{x, \text{bulk}'}$ is a copy of $\mathcal{H}_{x, \text{bulk}}$,
\begin{equation}
    V_{\text{TN}} = \left( \otimes_{x \in \mathcal{V}} \langle \text{MAX} |_{x} \right) | \Psi_{\text{TN}} \rangle \: , \qquad V_{\text{TN}} : \otimes_{x} \mathcal{H}_{x, \text{bulk}'} \rightarrow \otimes_{x} \mathcal{H}_{x, \partial} \: .
\end{equation}
Figure \ref{fig:generic_TN} provides a pictorial representation for the construction of $V_\text{TN}$ from $\mathcal{G}$.

\subsection{Tensor network model for closed universes}

\begin{figure}
\centering

\begin{subfigure}{0.35\textwidth}
  \includegraphics[width = \textwidth]{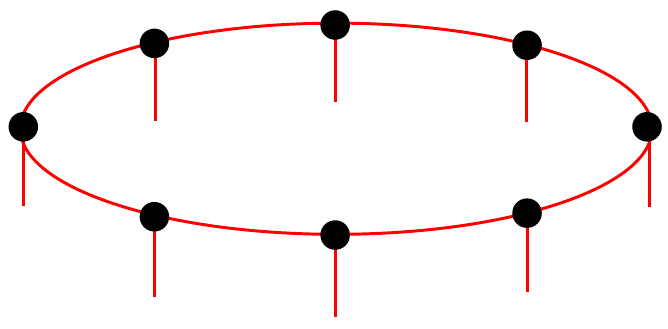}
  \caption{}
  \label{subfig:CU_TN_1}
\end{subfigure} \hfill
\begin{subfigure}{0.6\textwidth}
  \centering
  \input{tikz/CU_TN_2}
  \caption{}
  \label{subfig:CU_TN_2}
\end{subfigure}\hfill

\caption{(a) An eight-node tensor network representing $V_{\text{CU}}$ for a spatial slice of a closed universe with topology $S^1$. (b) A circuit representation of $V_\text{CU}$ for the simpler case of a graph with two nodes, where $O_{1}, O_{2}$ are random orthogonal matrices. The input to the tensor network is a state in $\hs_\text{eff} = \hs_{b_{1}} \otimes \hs_{b_{2}}$, and the contraction of legs between nodes is captured by the post-selection onto $\langle \text{MAX}|$.}
\label{fig:CU_TN}

\end{figure}

The utility of such tensor network models has recently been understood to extend into the regime in which the bulk-to-boundary map is non-isometric \cite{akers_black_2024,bueller_tensor_2024}.
The most extreme such example occurs in the case of a tensor network representing the state of a closed universe. A simple tensor network which will be useful for our purposes, which we denote by $V_{\text{CU}}$, is shown in figure \ref{fig:CU_TN}. In the 2-node example considered by \cite{akers_observers_2025},\footnote{\label{fn:convention}Note that \cite{akers_observers_2025} used a different convention for their tensor network models, where the map $V_\text{CU}$ was defined as the Hermitian conjugate of (\ref{eq:contracted}).} $V_\text{CU}$ takes the explicit form
\begin{equation}
    V_\text{CU} = d_{b_1}d_{b_2}\sqrt{d_{e_1}d_{e_2}} \, \Big( \langle\text{MAX}|_{b_1 b_1'} \langle\text{MAX}|_{b_2 b_2'} \langle\text{MAX}|_{e_1 e_2} \Big) \Big( O_1 |0\rangle_{b_1 e_1} \otimes O_2 |0\rangle_{b_2 e_2} \Big)
\end{equation}
where prefactors have been included to preserve state normalization for average choices of $O_1$ and $O_2$. Since there are no boundary legs, the entire output of the tensors in the model is post-selected on, making this tensor network a linear map $V_{\text{CU}} : \mathcal{H}_{\text{eff}} \rightarrow \mathbb{C}$. This is intended to capture the one-dimensional nature of the closed universe Hilbert space in quantum gravity, as suggested by the ``statistical approach'' to the gravitational path integral (amongst other arguments). 

The argument based on the statistical approach to the gravitational path integral has its own avatar in the tensor network model \cite{harlow_quantum_2026, akers_observers_2025}, and it will be instructive to describe this, since it gives us a concrete handle on Hilbert space dimension when we move to the path integral setting. A helpful estimate of this dimension can be found by computing the variance $\sigma^2$ of the inner product defined in (\ref{eq:variation}) between two bulk states $|\psi\rangle$ and $|\phi\rangle$ with respect to an ensemble of holographic maps obtained by taking the tensors $O_{1}, O_{2}$ to be independent Haar random orthogonal matrices. One has upon averaging
\begin{equation} \label{eq:first_inner_TN}
    \overline{\langle \phi_{1} |_{b_{1}} \langle \phi_{2} |_{b_{2}} V_{\text{CU}}^{\dagger} V_{\text{CU}} | \psi_{1} \rangle_{b_{1}} | \psi_{2} \rangle_{b_{2}}} = \langle \phi_{1} | \psi_{1} \rangle_{b_{1}} \langle \phi_{2} | \psi_{2} \rangle_{b_{2}}
\end{equation}
and
\begin{multline} \label{eq:second_inner_TN}
    \overline{|\langle \phi_{1} |_{b_{1}} \langle \phi_{2} |_{b_{2}} V_{\text{CU}}^{\dagger} V_{\text{CU}} | \psi_{1} \rangle_{b_{1}} | \psi_{2} \rangle_{b_{2}}|^{2}} \\
    = | \langle \phi_{1} |  \psi_{1} \rangle_{b_{1}} |^{2} | \langle \phi_{2} |  \psi_{2} \rangle_{b_{2}} |^{2} + 1 + | \langle \phi_{1}^{*} | \psi_{1} \rangle_{b_{1}}|^{2} | \langle \phi_{2}^{*} | \psi_{2} \rangle_{b_{2}}|^{2} + \mathcal{O}(1/d_{e}) \: ,
\end{multline}
where $d_{e}$ is the dimension of the Hilbert space for the contracted legs. By subtracting the square of (\ref{eq:first_inner_TN}) from (\ref{eq:second_inner_TN}), we find that the variance is $\mathcal{O}(1)$. This is a signature of the one-dimensional Hilbert space; if we think of a fixed holographic map as mapping orthogonal vectors in $\mathcal{H}_{\text{eff}}$ to random vectors in $\mathcal{H}_{\text{fun}}$, we anticipate that the dimension of $\mathcal{H}_{\text{fun}}$ should be approximately 
\begin{equation}
    \dim \hs_\text{fun} \sim \frac{1}{\sigma^2} \: .
\end{equation}
Therefore, an $\mathcal{O}(1)$ variance implies a fundamental Hilbert space of dimension $\mathcal{O}(1)$.

\subsection{Introducing an observer}

\begin{figure}
    \centering
    \input{tikz/Obs_2}
    \caption{A circuit representation of $V_{\text{ob}}$ obtained from the closed universe tensor network in figure \ref{subfig:CU_TN_2} by identifying the degrees of freedom in $b_{1}$ with an observer. }
    \label{fig:CU_ob_2}
\end{figure}
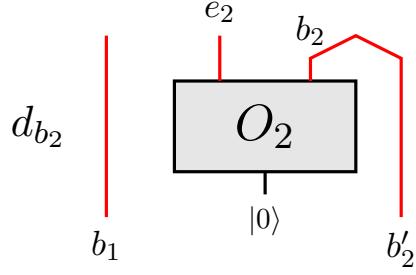

\begin{figure}
    \centering
    \includegraphics[width=0.9\linewidth]{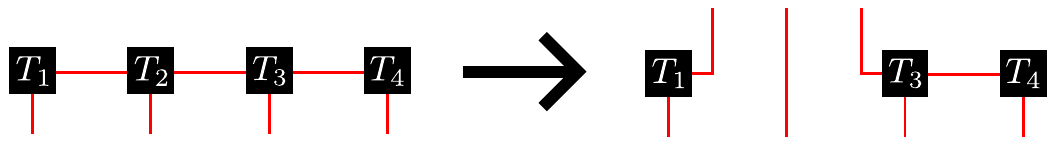}
    \caption{Pictorial representation of the CO rules for including an observer in a tensor network for a closed universe. Here, we take the observer region $b$ to correspond to degrees of freedom at node 2 in the graph; $V_\text{ob}$ is obtained by removing the tensor at the location of the observer in $V_\text{CU}$, leaving dangling legs at the observer's boundary and the observer's bulk effective degrees of freedom.}
    \label{fig:CU_ob}
\end{figure}

The proposed CO rules \cite{akers_observers_2025} suggest that once a subset $a\subset \{b_i\}$ of the effective degrees of freedom has been associated with an observer, one can modify the holographic map $V_{\text{CU}}$ to produce a new holographic map $V_{\text{ob}}$ from the bulk Hilbert space to a non-trivial Hilbert space associated with the observer. The ingredients of this transformation $F_{\text{ob}}(V_{\text{CU}}) = V_{\text{ob}}$ are:
\begin{itemize}
    \item Remove any tensors acting on the observer degrees of freedom in $a$ (including any post-selection or ancilla used to define the tensor). 
    \item Normalize so that the resulting tensor network maps unit vectors to unit vectors (at least on average if we have an ensemble). 
\end{itemize} 
This is shown in figure \ref{fig:CU_ob_2} for the 2-node model of figure \ref{subfig:CU_TN_2}, choosing the observer to correspond to degrees of freedom in $b_{1}$.
For more generic tensor networks
\begin{equation} \label{eq:before_hollow}
    V_{\text{CU}} = \left( \otimes_{x \in \mathcal{V}} \langle \text{MAX}|_{x} \right) \left( \otimes_{(xy) \in \mathcal{E}} \langle \text{MAX}|_{xy} \right) \left( \otimes_{x \in \mathcal{V}} | T_{x} \rangle \right) \: ,
\end{equation}
then $V_{\text{ob}}$ is given by\footnote{In the conventions used by \cite{akers_observers_2025} (see footnote \ref{fn:convention}), $V_\text{ob}$ does not remove $|\text{MAX}\rangle_{xy}$ on the boundary of the observer $\partial a$ since they are neither post-selection or ancilla. In either convention, these $\partial a$ degrees of freedom remain an output; in (\ref{eq:after_hollow}) they are a part of $|T_x\rangle$, while in \cite{akers_observers_2025} they are a part of $|\text{MAX}\rangle_{\partial a}$.}
\begin{equation} \label{eq:after_hollow}
    V_{\text{ob}} = \left( \otimes_{x \in \mathcal{V}} \langle \text{MAX}|_{x} \right) \left( \otimes_{(xy) \in \mathcal{E} \setminus \mathcal{E}_{a} \setminus \partial a} \langle \text{MAX}|_{xy} \right) \left( \otimes_{x \in \mathcal{V} \setminus a} | T_{x} \rangle \right) \: ,
\end{equation}
where $\mathcal{E}_{a}$ denotes edges in the interior of $a$ and $\partial a$ denotes edges on the boundary of $a$. 
This is shown in figure \ref{fig:CU_ob}.

It is immediately evident that the fundamental Hilbert space associated with $V_{\text{ob}}$ is no longer generically trivial, since the network no longer post-selects on all outputs. The fundamental Hilbert space has dimension at least $d_\text{ob}$, and at most $d_\text{ob} e^{\mathcal{A}_\text{ob}/4G_N}$, where $d_\text{ob}$ is the dimension of the effective field theory subspace associated with the observer, and $\mathcal{A}_\text{ob}/4G_N$ is the log dimension of the Hilbert space associated with the edges of the tensor network connecting the observer's subregion to its complement. The precise value of the dimension would depend on details of the tensor network.

The identification of the observer modifies the calculation of statistical moments of overlaps of states in the fundamental Hilbert space. For the model in figure \ref{fig:CU_ob_2}, one finds \cite{akers_observers_2025}
\begin{equation}
    \overline{\langle \phi |_{b_{1}} \langle \phi ' |_{b_{2}} V^{\dagger}_{\text{ob}} V_{\text{ob}} | \psi \rangle_{b_{1}} | \psi ' \rangle_{b_{2}}} = \langle \phi | \psi \rangle_{b_{1}} \langle \phi ' | \psi ' \rangle_{b_{2}}  
\end{equation}
and
\begin{multline} \label{eq:inner2_obsTN}
    \overline{| \langle \phi |_{b_{1}} \langle \phi ' |_{b_{2}} V^{\dagger}_{\text{ob}} V_{\text{ob}} | \psi \rangle_{b_{1}} | \psi ' \rangle_{b_{2}}|^{2}} \\
    = | \langle \phi | \psi \rangle_{b_{1}}|^{2} \left( |  \langle \phi ' | \psi ' \rangle_{b_{2}} |^{2} + \frac{1}{d_{e}} \left( 1 + |\langle (\phi')^{*} | \psi ' \rangle_{b_{2}}|^{2} \right) \right) + \mathcal{O}(1/d_{e} d_{b_{2}}) \: ,
\end{multline}
so the variance is suppressed by $\mathcal{O}(1/d_{e})$, the dimension of the contracted edge. This suggests a fundamental Hilbert space with dimension $\mathcal{O}(d_{e})$, which is interpreted as an area in a one-dimensional network.

\subsection{Equivalence to the hollowing map}

Our main observation in this subsection is that the ``observer-promotion map'' $F_{\text{ob}}(V_{\text{CU}}) = V_{\text{ob}}$ is precisely the same as the ``hollowing map'' used by KRR to promote a subsystem of the bulk effective theory to a subsystem of the fundamental Hilbert space to which a density matrix can be associated. Acting on $V_\text{CU}$ in equation (\ref{eq:before_hollow}), KRR's hollowing map gives precisely $V_\text{ob}$ in (\ref{eq:after_hollow}).

\begin{figure}
    \centering
    \includegraphics[width=0.9\linewidth]{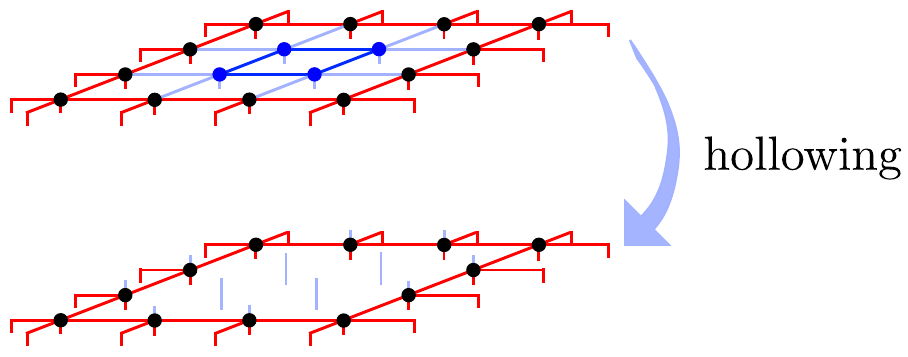}
    \caption{Pictorial representation of KRR's hollowing map $V_{a}$. Tensors located within some bulk subregion $a$ (blue nodes) are removed, creating open legs from previously in-plane legs on $\partial a$ and bulk legs on $a$.}
    \label{fig:hollowing_TN}
\end{figure}

The authors of \cite{kaya_hollowgrams_2025} are interested in starting with some tensor network state, and calculating the entropy associated with some promoted subset $a$ of bulk degrees of freedom. To do so, they apply the hollowing map to produce a modified tensor network $V_{a}$, thereby producing new uncontracted legs in the network; see figure \ref{fig:hollowing_TN}. The density matrix $\rho_{a}$ on these uncontracted legs whose entropy we would like to determine can be obtained by tracing over the original boundary legs, and the R{\'e}nyi entropy can then be computed by gluing together $n$ such copies. KRR demonstrate that, upon using the random tensor network techniques of \cite{Hayden:2016cfa}, this leads to the result
\begin{equation}
    S(\rho_{a}) = S_{\text{gen}}\left[ E(a) \right] \: ,
\end{equation}
where $E(a)$ is the generalized entanglement wedge \cite{bousso_entanglement_2023, bousso_holograms_2023} associated with $a$ (bounded by the minimal generalized entropy cut through the network in the exterior of $a$ which is homologous to $\partial a$). We recall that the generalized entropy in the tensor network context is 
\begin{equation}
    S_{\text{gen}}\left[ a \right] = \ln d_{\gamma_{a}} + S(\rho_{a}^{\text{bulk}}) \: ,
\end{equation}
where $\gamma_{a}$ is the cut through the network separating region $a$ from its complement, $d_{\gamma_{a}}$ is the Hilbert space dimension associated with this cut, and $\rho_{a}^{\text{bulk}}$ is the bulk state on the region $a$.

By way of illustration, we can consider how this works in more detail when applying the KRR method to the tensor network for a closed universe.
We can consider the 2-node model of \cite{akers_observers_2025} depicted in figures \ref{subfig:CU_TN_2} and \ref{fig:CU_ob_2}; we will relabel $b_{1} \rightarrow \text{ob}$ and $b_{2} \rightarrow M$, identifying these with some observer and matter subsystems in the effective description. As before, $O_{1}, O_{2}$ are taken to be Haar random orthogonal matrices. 
If we take the input state to factorize as $\rho_\text{ob} \otimes \rho_{M}$ in the effective description, then the state in the fundamental Hilbert space is
\begin{align*}
    \rho_{a} &= V_{a} (\rho_\text{ob} \otimes \rho_M) V_{a}^\dagger \\
    &= d_{M} \, \rho_\text{ob} \otimes \langle\text{MAX}|_{M} O_2 \big( \rho_M \otimes |0\rangle\langle0| \big) O_2^\intercal |\text{MAX}\rangle_M \in \hs_\text{ob} \otimes \hs_{e_2} \:, \numberthis
\end{align*}
where $V_a$ is constructed by hollowing $O_1$ from the tensor network. One then has
\begin{equation}
    \overline{\text{tr}(\rho_{a}^{n}) } = \tr (\rho_\text{ob}^n) \left( \frac{1}{d_e^{n-1}} + \tr (\rho_M^n) + \dots \right)
\end{equation}
where we have included the leading order contributions from contractions in the Haar integration of $O_2$ (within each replica and cyclically, respectively).
 
The annealed entropy is then given by
\begin{equation}
    S_{\text{ann}}(\rho_{a}) = S(\rho_\text{ob}) + \lim_{n\to1} \frac{1}{1-n} \ln \left( \frac{1}{d_e^{n-1}} + \tr (\rho_M^n) + \dots \right)
\end{equation}
In particular, 
\begin{equation}
    S_{\text{ann}}(\rho_{a}) = \begin{cases}
        S(\rho_{\text{ob}}) + S(\rho_{M}) & \ln(d_e) \gg S(\rho_{M}) \\
        S(\rho_{\text{ob}}) + \ln(d_{e}) & \ln(d_e) \ll S(\rho_{M})
    \end{cases} \: .
\end{equation}
We recognize this as recovering the usual behaviour of the generalized entropy prescription in the large $d_e$ limit. When the geometric contribution to the generalized entropy dominates the matter contribution, then the RT surface will be the trivial (empty) surface, the generalized entanglement wedge is the entire universe, and the entropy picks up a contribution from the bulk matter. On the other hand, when the matter contribution dominates the geometric contribution, the RT surface will be at the boundary of the observer's region, thus incurring a geometric contribution $\ln(d_{e})$ but avoiding a bulk entropy contribution. Note that when the bulk effective state is pure, $S(\rho_\text{ob}) = S(\rho_M) = 0$, we find $S_\text{ann}(\rho_a) = 0$.

\section{KRR path integral rules for bulk observers} \label{sec:GPI}

Having identified the equivalence between the CO and KRR tensor network rules, let us now turn to applying the KRR prescription for the gravitational path integral to bulk observers. We begin with a brief review of the fixed geometry method of calculating holographic R\'{e}nyi entropies \cite{akers_holographic_2019,dong_flat_2019,dong_modified_2024}, first for a boundary subregion and then for KRR's generalization to a bulk subregion. We refer the reader to sections 3 and 4 of \cite{kaya_hollowgrams_2025} for more details. In section \ref{sec:KRR_CU}, we apply KRR's rules to the generalized entropy of a closed universe observer.

\subsubsection*{Fixed geometry states}

Replica calculations of entropies in random tensor networks \cite{Hayden:2016cfa} have a very close analogue in a class of states known as fixed geometry states \cite{akers_holographic_2019,dong_flat_2019}. It is therefore unsurprising that extrapolating KRR's tensor network rules to gravity will require working with fixed geometry states. 

For our purposes, fixed geometry states will be understood as essentially the analogue of position eigenstates for the bulk metric operator.\footnote{More accurately, it is typically assumed that the wavefunction for $| h \rangle$ is sharply peaked in the semi-classical limit, but still broad enough to contain many metric eigenstates.} We will denote them by $| h \rangle$, with $h$ understood to represent a particular configuration of the metric (and other matter fields if present). Their inner products may be computed via a path integral with fixed geometry initial and final boundary conditions:
\begin{align*}
    \langle h'|h\rangle &= 
    \vcenter{\hbox{
    \input{tikz/h_inner}
    }} + \dots \\
        &= \delta(h - h') + \dots \numberthis
\end{align*}
In situations with asymptotic boundaries, like the two-sided case shown, we anticipate that these states are perturbatively orthogonal; see \cite{iliesiu_nonperturbative_2024} for non-perturbative corrections in the context of JT gravity. 

Inserting a resolution of the identity consisting of fixed geometry states is a helpful tool in replica calculations, because identifying the dominant saddle in the path integral is simplified within each fixed geometry sector. Indeed, suppose that we have some general time reflection-symmetric state $|\psi \rangle$ defined by some Euclidean path integral, and we would like to compute the entropy associated with the reduced density matrix
\begin{equation}
    \rho_{A} \equiv \frac{\text{tr}_{\bar{A}}( | \psi \rangle \langle \psi | )}{\langle \psi | \psi \rangle} \: ,
\end{equation}
where $A$ is some boundary spatial subregion and $\bar{A}$ is its spatial complement. Applying the familiar replica trick, we proceed by calculating $\text{tr}(\rho_{A}^{n})$. Inserting $n$ resolutions of the identity involving fixed geometry states, we obtain $\text{tr}(\rho_{A}^{n}) = \frac{Z_{n}}{Z_{1}^{n}}$ with
\begin{equation}
    Z_{n} = \sum_{h_{1}, \ldots, h_{n}} \sum_{h_{1}', \ldots, h_{n}'} \left( \prod_{k=1}^{n} \langle h_{k} ' | \psi \rangle \langle \psi | h_{k} \rangle \right) \text{tr} \left[ \prod_{\ell=1}^{n} \text{tr}_{\bar{A}}( | h_{\ell} ' \rangle \langle h_{\ell} |) \right] \: .
\end{equation}
where $k$ and $\ell$ index the $n$ replicas.

The first advantageous observation is that the dependence on $| \psi \rangle$ has factored out of the trace, contributing a prefactor which is a product of terms $\psi(h_{k}), \psi^{*}(h_{k}')$ which can be estimated by a gravitational saddle point calculation by imposing $h$ as a boundary condition on the time-symmetric slice $\Sigma$:
\begin{equation} \label{eq:fixedgeo_bdys}
    \psi(h) = 
    \vcenter{\hbox{
    \input{tikz/psi_h}
    }}
    \qquad
    |\psi(h)|^2 = 
    \vcenter{\hbox{
    \input{tikz/psi_h_2}
    }}
\end{equation}
It is then standard to make a ``diagonal approximation'' and assume that the calculation is dominated by a single diagonal sum $h_{k} = h_{k} ' = h$. This is connected to the assumption of replica symmetry, given that off-diagonal contributions to the sum necessarily break this symmetry, but it does not require replica symmetry; see \cite{penington_diagonal_2024, Held:2024qcl} for further details. Under this assumption, $Z_n$ simplifies to
\begin{equation} \label{eq:rho_B}
    Z_{n} = \sum_{h} | \psi (h) |^{2n} \: \text{tr} \left[ \rho_{A}(h)^{n} \right] \: , \qquad \rho_{A}(h) \equiv \text{tr}_{\bar{A}}( | h \rangle \langle h |) \: .
\end{equation}

It remains to compute the trace of $\rho_{A}(h)^{n}$, an object which is completely independent of $| \psi \rangle$ and depends only on the properties of the fixed geometry states. The dominant contribution to this trace comes from the degenerate geometry where the $n$ bra/ket replicas are glued according to the identity permutation in a domain of the fixed geometry slice with boundary including $\bar{A}$, and according to the cyclic permutation $(123\ldots n)$ in a domain with boundary including $A$. For example, in the case of $n=2$ with $A$ given by the right boundary, 
\begin{equation} \label{eq:saddle_renyi2}
    \tr_A\Big((\tr_{\bar{A}} |h\rangle\langle h|)^2\Big) = 
    \vcenter{\hbox{
    \input{tikz/tr_rho2_h}
    }} + \quad \dots
\end{equation}
In this configuration, the on-shell action receives a contribution $(n-1)\mathcal{A}/4G_N$ from the area of the domain wall separating the two permutation domains, which is a conical defect due to the gluing. When matter fields are incorporated, there will generally also be contributions from inner products of matter states inside the two domains, with the pattern of contractions between states in different replicas determined by the relevant permutation. One thus arrives at the expression
\begin{equation}
    Z_{n} = \sum_{h} | \psi(h) |^{2n} e^{-(n-1) \mathcal{A} [\gamma_{A}(h)]/4G_N} \text{tr}\left[ (\rho^{\text{bulk}}_{E(A)}(h))^{n} \right] \: ,
\end{equation}
where $\gamma_{A}(h)$ is a surface homologous to $A$ in geometry $h$, $E(A)$ is the associated homology region, and $\rho_{E(A)}^{\text{bulk}}(h)$ is the density matrix of bulk matter in that region. In particular, the surface $\gamma_{A}(h)$ is the one that maximizes the expression in the sum; for example, it is the minimal area surface in the semi-classical limit $G_{N} \rightarrow 0$. Performing a saddle-point approximation, we may assume that this sum is dominated by some particular geometry $h_{n}$. The replica trick then gives
\begin{equation}
    \begin{split}
        S(\rho_{A}) & = \lim_{n \rightarrow 1} \frac{1}{1-n} \ln \left( Z_{n} / Z_{1}^{n} \right) \\
        & = \lim_{n \rightarrow 1} \frac{1}{1-n} \left[ 2n \ln \Bigg| \frac{\psi(h_{n})}{\psi(h_{1})} \Bigg| - (n-1) \frac{\mathcal{A}[\gamma_{A}(h_{n})]}{4G_N} + \ln \left( \text{tr}\left[(\rho_{E(A)}^{\text{bulk}}(h_{n}))^{n}\right] \right) \right] \: .
    \end{split}
\end{equation}
Assuming that $h_{n} \rightarrow h_{1}$ in the limit, one recovers the Ryu-Takayanagi prescription
\begin{equation}
    S(\rho_{A}) = \frac{\mathcal{A}[\gamma_{A}(h_{1})]}{4G_N} + S(\rho_{E(A)}^{\text{bulk}}(h_{1})) \: .
\end{equation}

Note that this procedure very closely mirrors the calculation in random tensor networks \cite{Hayden:2016cfa}, which can be reduced to finding a domain wall configuration which minimizes the free energy associated with an Ising model with suitable boundary conditions. This similarity reflects the close relationship between fixed geometry states in gravity and tensor network states.

\subsubsection*{The KRR hollowing map}

KRR sought to generalize the above fixed geometry prescription for computing generalized entropies to a bulk subregion $a\subset\Sigma$.\footnote{The specification of $a$ must be gauge-invariant. We will comment on methods for gauge-invariantly defining $a$, including in the case of a closed universe, in section \ref{sec:gauge-inv} and continue with a generically defined subregion here.} KRR's primary observation was that the same procedure can be used to recover the BP proposal \cite{bousso_entanglement_2023,bousso_holograms_2023} if one allows $a$ to play the role of $A$ above by enforcing a boundary condition for the allowed gluings within $a$.

To formalize this, KRR drew inspiration from the tensor network produced by the hollowing map, $V_a = F_{a}(V_{\text{TN}})$. In gravity, they formally defined an analogous map on a basis of fixed geometry states,\footnote{In the tensor network case, $V_{a}$ was defined as a map $V_{a} : \mathcal{H}_{\text{eff}} \rightarrow \mathcal{H}_{\text{fun}, a}$. In gravity, the analogous object is a map $V_{a, h}$ for a fixed geometry $h$, which is a map acting on the EFT Hilbert space associated with this geometry. The full $V_{a}$ in gravity is then a direct sum over $h$ of $V_{a, h}$. Note also that we are abusing notation by now using $| h \rangle$ to refer to a state in the perturbative bulk Hilbert space.}
\begin{equation}
    V_{a} |h  \rangle = | h_{a} \rangle \: .
\end{equation}
This map is implicitly defined via a modification of the path integral rules associated with the fixed geometry states, which places the bulk subregion $a$ on the same footing as the spatial boundary $\partial\Sigma$, thereby modifying e.g.\ replica calculations:
\begin{equation} \label{eq:V_a}
    V_a: 
    \vcenter{\hbox{
    \input{tikz/h_before_after}
    }}
\end{equation}
where blue denotes the portions of $\Sigma$ where boundary conditions are imposed in the path integral. We emphasize that this does not alter the fixed geometry and field configuration data $h$, only the boundary conditions that must be satisfied in path integral manipulations involving these fixed geometry states, as we elucidate presently. 

Returning to (\ref{eq:rho_B}), KRR finds the generalized entropy of $a$ by replacing $\rho_A(h)$ with $\rho_a(h)$ defined as
\begin{equation} \label{eq:rho_a}
    \rho_a(h) = \tr_{\partial\Sigma} V_a|h\rangle\langle h|V_a^\dagger \: .
\end{equation}
The trace of $\rho_a(h)^n$ is then found by enforcing identity permutations on $\partial\Sigma$, cyclic permutations on $a$, and optimizing over choices of gluing in the remainder of the bulk $\Sigma\backslash a$. In the case of $n=2$, this schematically gives
\begin{equation} \label{eq:rho_a2}
    \tr\Big((\tr_{\partial\Sigma} V_a|h\rangle\langle h|V_a^\dagger)^2\Big) = 
    \vcenter{\hbox{
    \input{tikz/tr_rhoa2}
    }} + \dots
\end{equation}
where $a$ is denoted in thick purple. This leads to a new domain wall $\gamma_a(h)$ separating bulk gluings that is homologous to $\partial a$. Optimizing over the location of $\gamma_a(h)$ defines the generalized entanglement wedge $E(a)$ for $a$ and reproduces an RT-like formula for the generalized entropy of $a$:
\begin{equation} \label{eq:S(a)}
    S(\rho_{a}) = \frac{\mathcal{A}[\gamma_{a}(h_{1})]}{4G_N} + S(\rho_{E(a)}^{\text{bulk}}(h_{1})) \: .
\end{equation}
Had we not enforced the cyclic gluing of $a$ through the inclusion of $V_a$ in (\ref{eq:rho_a}), the dominant contribution to a replica calculation like (\ref{eq:rho_a2}) would be given by an identity permutation on the entire bulk so that 
\begin{equation}
    Z_n = Z_1^n \qquad \implies \qquad \tr (\rho_a^n) = 1 \qquad \implies \qquad S(\rho_a) = 0.
\end{equation}
Of course, there is no generalized entanglement wedge in this case.

Defining $V_a$ as in (\ref{eq:V_a}) suggests that the holographic dual of $V_a|h\rangle$ should be interpreted as living in a larger Hilbert space, one that includes both the boundary $\partial\Sigma$ and the bulk subregion $a$. This is reminiscent of the assumption motivating the CO tensor network rules for observers: \textit{the observer is already a part of the fundamental description}. 
It is also a manifestation of AAIL's proposal that the observer's subregion (e.g.\ their worldline) should be on the same footing as a more conventional holographic boundary. Inspired by this, we will use the KRR hollowing map $V_a$ as a new way to incorporate observers into path integrals by defining $a$ as the subregion occupied by some observer. In the next subsection, we apply these techniques to an observer in a closed universe.

\subsection{KRR for an observer in a closed universe} 
\label{sec:KRR_CU}

We now apply KRR's hollowing prescription to an extended observer occupying a subregion of a closed universe. Here we will compute the generalized entropy of the observer; in section \ref{sec:bulk_HS} we will use KRR's prescription to estimate the dimension of the bulk Hilbert space. 

KRR's method for computing the generalized entropy of a bulk subregion reviewed above can be applied directly to closed universes. We will consider a state $| \psi \rangle$ for the closed universe, specified by an asymptotic boundary condition (possibly including matter sources), and decompose it in a basis of fixed geometry states on $\Sigma$:\footnote{
At the non-perturbative level, this decomposition is no longer possible -- the fundamental Hilbert space is one-dimensional, and all fixed geometry states are non-perturbatively equivalent. However, by combining perturbative calculations with the hollowed encoding map $V_{a}$ on perturbative states, we can produce the anticipated result for the observer's entropy, and we can infer that this is a good approximation to a putative non-perturbative calculation. This was also the approach to the derivations of the semi-classical RT/BP formulas above. From the perspective of the AAIL formalism, this is similar to the observation that one can already see that the variance of the inner product becomes small via a perturbative calculation, without actually needing to resum the full genus expansion. 
}
\begin{equation}
    \psi(h) = 
    \vcenter{\hbox{
    \input{tikz/closed_state}
    }} \: .
\end{equation} 
Just as in the case of a universe with boundary, we define a ``hollowed'' state by (gauge-invariantly) specifying some subregion $\text{ob}\subset\Sigma$ occupied by an observer,
\begin{equation} \label{eq:hollowed_obs}
    \psi(h_\text{ob}) = 
    \vcenter{\hbox{
    \input{tikz/closed_Vobs}
    }}
\end{equation}
which allows us to impose boundary conditions on the observer's subregion. To compute the generalized entropy of the observer, we again replace $\rho_B(h)$ in (\ref{eq:rho_B}) with $\rho_\text{ob}(h)$ defined here as
\begin{equation}
    \rho_\text{ob}(h) = V_\text{ob}|h\rangle\langle h| V_\text{ob}^\dagger \: .
\end{equation}
Note there is no partial trace over $\partial\Sigma$ like that appearing in (\ref{eq:rho_a}) since $\partial\Sigma$ is empty here. The only boundary condition that we can implement on the gluing of $n$ copies of $\Sigma$ is the cyclic permutation on the observer's subregion; the rest of $\Sigma\backslash \text{ob}$ is allowed to be glued in all possible ways. In the case of $n=2$, we can consider three types of gluings for $\Sigma\backslash \text{ob}$:
\begin{equation} \label{eq:three_terms} 
    \vcenter{\hbox{
    \input{tikz/tr_rho2_closed}
    }}
\end{equation}
The first two terms correspond to the usual identity and cyclic permutations of $\Sigma\backslash \text{ob}$ familiar from universes with boundary, while the third term is a new gluing permitted by gauging CRT invariance in closed universes \cite{harlow_gauging_2023}. The second term dominates the sum since it lacks a conical defect; in this case,
\begin{equation}
    Z_n = Z_1^n \qquad \implies \qquad \tr \rho_\text{ob}^n = 1 \qquad \implies \qquad S(\rho_\text{ob}) = 0 \: .
\end{equation}
In fact the other two terms are no different; if we naively follow KRR's prescription with just the first term in (\ref{eq:three_terms}), we find an RT-like formula for the generalized entropy of the observer:
\begin{equation} \label{eq:S_obs}
    S(\rho_\text{ob}) \approx \frac{\mathcal{A}[\partial E(\text{ob})]}{4G} + S(\rho_{E(\text{ob})}^{\text{bulk}}(h)) \: .
\end{equation}
However, the lack of boundary means the minimal surface homologous to the observer's boundary is the empty surface, $\gamma_\text{ob} = \varnothing$, again giving $S(\rho_\text{ob}) = 0$. This result is not surprising and is interpreted as a consequence of the one-dimensional Hilbert space. 

We note that it is possible to make the observer's generalized entropy non-zero by adding matter to the observer's environment $\Sigma\backslash \text{ob}$ in a mixed state.\footnote{Alternatively, one could add a second observer to the closed universe. Computing the generalized entropy of just one observer corresponds to enforcing a cyclic gluing (purple) on their subregion and identity gluing (blue) on the other. The dominant gluing would then take the form of the first term in (\ref{eq:three_terms}), leading to a non-zero generalized entropy. We thank Takato Mori and Pratik Rath for discussions on this alternative.} We could imagine adding an external reference to purify this matter, in which case $\rho_\text{ob}(h)$ involves a partial trace over $R$,
\begin{equation}
    \rho_\text{ob}(h) = \tr_R ( V_\text{ob} |h\rangle\langle h| V_\text{ob}^\dagger ).
\end{equation}
Now the second term in (\ref{eq:three_terms}) will be suppressed by $\tr \rho_R^2$, allowing the other two to compete. A non-trivial homologous surface can be made to dominate by increasing the entanglement with $R$, permitting a non-zero generalized entropy and entanglement wedge $E(\text{ob}) \subsetneq \Sigma$ for the observer. We provide an explicit example of this in the context of a JT gravity toy model in section \ref{sec:Sgen_obs_JT}. 

Before continuing to this example, it is important to return to the question of gauge-invariantly specifying the observer's subregion, which we discuss presently. 

\subsubsection{Gauge-invariant subregion}
\label{sec:gauge-inv}

Whether or not the universe is closed, the subregion $a$ occupied by the observer must be defined in a gauge invariant way. In their work, KRR discuss a few methods for doing this in a universe with boundary; these include $(i)$ fixing the proper distance from an asymptotic boundary and $(ii)$ fixing geometrical data at the corner of the bulk subregion, such as the dilaton value in JT gravity.

Of course, option $(i)$ is useless in a closed universe, as there is no asymptotic boundary to which we may dress a subregion. This leaves option $(ii)$, which is subtle in its own right. Consider the classical dilaton solutions in closed universe JT gravity found by Usatyuk, Wang, and Zhao (UWZ) \cite{usatyuk_closed_2024}. Unless matter insertions are included, the dilaton solutions are flat, in which case there is no special value to use for dressing. UWZ finds non-trivial dilaton solutions when matter insertions are included, and we could try dressing to either the matter insertion itself or a specified dilaton value in its now non-trivial profile.

\begin{figure}
    \centering
    \includegraphics[width=0.7\linewidth]{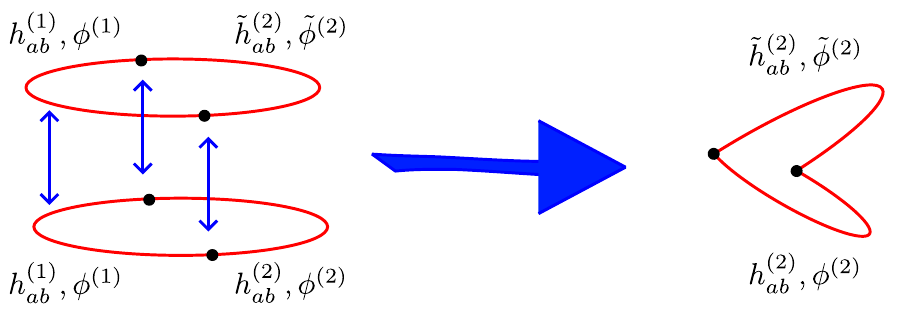}
    \caption{Proposal for path integral boundary conditions computing the overlap of states in the Hilbert space relative to an observer.}
    \label{fig:gluing}
\end{figure}

In principle, we can use the Euclidean path integral to prepare such states with asymptotic matter insertions at $\tau\to-\infty$. However, we have to confront the possibility that in some branches of the wavefunction the asymptotic insertion may become a very delocalized excitation on some prescribed slice. In such cases, dressing to a localized excitation or special dilaton value might not be possible.
Perhaps the simplest way to avoid this problem would be to work directly with fixed geometry states rather than asymptotic states, demanding fixed values for the geometry and fields in the observer's region in these states; the resulting path integral boundary conditions for the relative Hilbert space are shown in figure \ref{fig:gluing}. 

As an alternative, we could create a localized matter excitation using an asymptotic insertion and assume that it must remain localized throughout the path integral preparation of the bulk state. This is akin to AAIL's original assumption that an observer cannot be fluctuated away. We emphasize that this would be a modification to the definition of the non-perturbative path integral, but it might be justifiable if the computation is dominated by contributions where this assumption holds. 

For simplicity, we will work with AAIL's condition -- assuming a matter excitation remains localized -- in the following examples, leaving exploration of other options to future work.

\subsection{A toy example: JT gravity with probe matter} \label{sec:Sgen_obs_JT}

We illustrate the above analysis with a concrete example by considering closed universes in JT gravity coupled to matter. Setting $8\pi G = 1$, the Euclidean action is given by
\begin{equation}\label{eq:jt-action}
    I_{JT} = -S_0 \chi(\mathcal{M})-\dfrac{1}{2}\int_{\mathcal{M}}\sqrt{g}\Phi(R+2) -\int_{\partial\mathcal{M}}\sqrt{h}\Phi_b(K-1) + I_m.
\end{equation}
where $\Phi_0 = S_0/2\pi$ is a constant shift in the dilaton field $\Phi$ and $\chi$ is the Euler characteristic of the manifold $\mathcal{M}$.

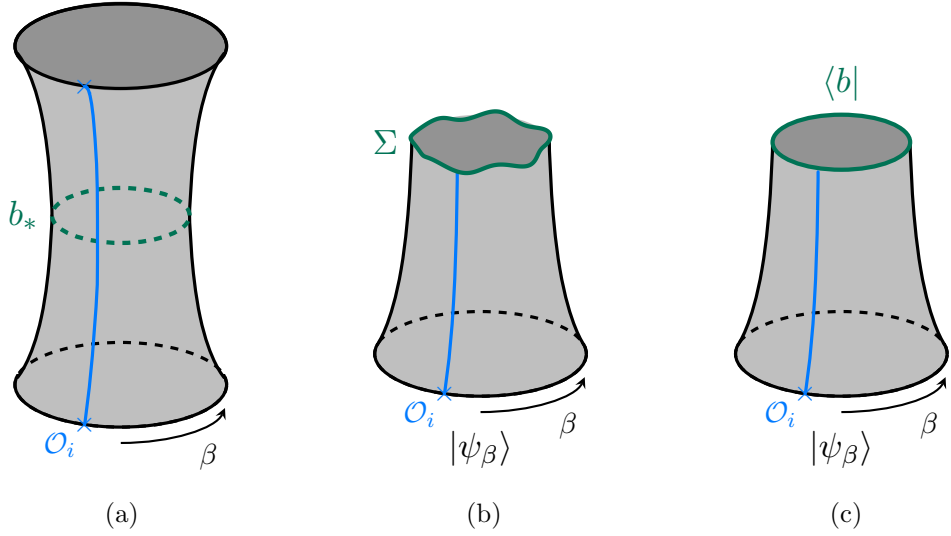
\begin{figure}
    \centering
    \begin{subfigure}[b]{0.30\textwidth}
        \centering
        \input{tikz/doubletrumpet-jt}
        \subcaption[]{}
    \end{subfigure}
    \begin{subfigure}[b]{0.30\textwidth}
        \centering
        \input{tikz/psi_beta}
        \subcaption[]{}
    \end{subfigure}
    \begin{subfigure}[b]{0.30\textwidth}
        \centering
        \input{tikz/psi_b_}
        \subcaption[]{}
    \end{subfigure}
    
    \caption{(a) Double trumpet geometry for a closed JT universe + matter. This is the saddle point solution to $\text{tr}(e^{-\beta H} \mathcal{O}_i)^2$ (b) The state $|\psi_\beta\rangle$ is defined by cutting open the double trumpet at the maximally symmetric worldslice $\Sigma$. (c) $|\psi_\beta\rangle$ can be written in the $|b\rangle$ basis by fixing $\Sigma$ to be the minimal worldslice of the double trumpet geometry with geodesic length $b$.}
    \label{fig:double-trumpet}
\end{figure}

We begin by considering a state $|\psi_\beta\rangle$ defined by cutting the Euclidean gravitational path integral computing the quantity $\text{tr}(e^{-\beta H} \mathcal{O}_i)^{2}$ at the time-reflection symmetric worldslice $\Sigma$. This setup was extensively studied in \cite{usatyuk_closed_2024}. Here, $\beta$ is the length of either Euclidean asymptotic boundary, $\mathcal{O}_i$ is a matter insertion operator, and $H$ is the Hamiltonian that is dual to JT gravity + matter. The saddle point solution to this quantity, referred to as the double trumpet geometry, is depicted in figure \ref{fig:double-trumpet}a. The dashed blue line is the minimal world slice with geodesic length $b_*$. The metric and dilaton profile for this on-shell solution are given by
\begin{equation} \label{eq:jt-solutions-trumpet}
    ds^2=d\rho^2 + b_*^2 \cosh(\rho)d\sigma^2, \hspace{10pt} \Phi(\rho, \sigma) = \frac{\pi{\phi_r}}{\beta}\cosh(\rho)\cosh(b_*\sigma).
\end{equation}
where $\phi_r$ is the renormalized value of the dilaton on the Euclidean asymptotic boundaries. Here, $\rho\in(-\infty, \infty)$, and $\sigma\sim \sigma+1$ with $\sigma \in[-\frac{1}{2}, \frac{1}{2}]$. The particle inserted by $\mathcal{O}_i$ sits at position $\sigma = \pm \frac{1}{2}$ for all values of $\rho$.

We denote by $\{|b\rangle\}_{b}$ the basis of states on closed geodesic slices of lenth $b$. The perturbative inner product between two $b$ basis states is given by $\langle b |b'\rangle = \frac{1}{b}\delta(b-b')$ \cite{usatyuk_closed_2024,abdalla_gravitational_2025}. The state $|\psi_\beta\rangle$ can be decomposed in this basis as $|\psi_\beta\rangle = \sum_b b\,\psi_\beta(b)|b\rangle$ where the wave function $\psi_\beta(b) = \langle b|\psi_\beta\rangle$ is obtained from the Euclidean path with an asymptotic boundary of renormalized length $\beta$ with matter insertion $\mathcal{O}_i$, and the additional boundary condition that $\Sigma$ is fixed to be the minimal worldslice with geodesic length $b$; see figure \ref{fig:double-trumpet}c. 

Evaluating the on-shell JT action (\ref{eq:jt-action}) over the trumpet geometries, it was shown in \cite{usatyuk_closed_2024} that the wavefunction takes the form

\begin{multline}\label{eq:b-basis-wf}
    \notag \psi_\beta(b) = \exp\Bigg[ -m \log\Bigg( \dfrac{\beta/\phi_r}{\cosh(\frac{b}{2})\arctan(\sinh(\frac{b}{2}))}\Bigg)
    \\
    +\dfrac{2\phi_r}{\beta}\arctan^2 \left( \sinh \left( \frac{b}{2} \right) \right)
    -\dfrac{4\phi_r}{\beta}\sinh \left( \frac{b}{2} \right) \arctan \left( \sinh \left( \frac{b}{2} \right) \right) \Bigg] \numberthis
\end{multline}
where $m$ is the mass of the inserted particle. We elaborate in appendix \ref{app:fixed_geom} on why the $|b\rangle$ basis works as a fixed geometry basis for closed JT gravity. 

\begin{figure}[t]
    \centering
    \begin{subfigure}{0.49\linewidth}
        \centering
        \input{tikz/probediagram-smaller}
        \caption{ }
    \end{subfigure}
    \begin{subfigure}{0.49\linewidth}
        \centering
        
        \input{tikz/dilaton_profile}
        \caption{ }
    \end{subfigure}
    \caption{(a) The observer subregion $a \subset \Sigma$ and the associated entanglement wedge $E(a)$. The entanglement wedge is entirely determined by the light probe's position relative to the observer. (b) Dilaton profile at $\rho = 0$ with the observer's location $\sigma=\pm1/2$ (marked by $\times$) identified.}\label{fig:renyi-jt-calc}
\end{figure}
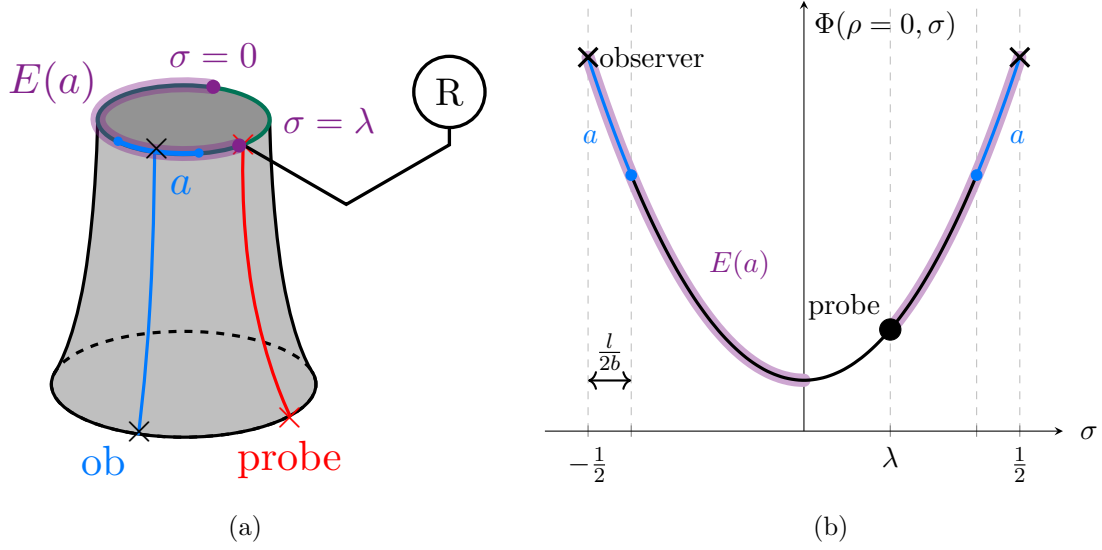

We are interested in computing the Rényi entropy of a subregion defined on $\Sigma$ using the KRR rules. To do so, we must first gauge-invariantly define such a region. Following the discussion in section \ref{sec:gauge-inv}, we will dress the region to the matter particle. One can do this in several ways; for instance, we choose to define $a\subset\Sigma$ by setting the proper distance between the matter particle and the interval's endpoints to be $\ell/2$.\footnote{For this region to be well-defined for arbitrary $b$, we will define it to be the entire slice $\Sigma$ when $b < \ell$.} See figure \ref{fig:renyi-jt-calc}b for an illustration. In this way, we interpret the matter particle and its associated subregion $a$ as the observer and the subregion they occupy. 

We first note that in the case where there is no other bulk matter, or the matter entropy is sub-leading, the Rényi entropy $S_n(a)$ vanishes for all values of $n$. This can be seen from the fact that the minimal homologous surface to $\partial a$ in this case is the empty surface, $\gamma_a=\varnothing$. The area functional hence vanishes, and we have
\begin{equation}
    Z_n = \sum_b |b\, \psi_\beta(b)|^{2n} \: .
\end{equation}
In this case, the saddle $b_n$ is the value $b$ that extremizes $|b\, \psi_\beta(b)|$ so that $b_n=b_1$ for all $n$, implying that $Z_n = Z_1^n$ and hence $S_n(a)=0$. In particular, the von Neumann entropy ($n\to 1$) vanishes, as expected.

In order to obtain non-trivial Rényi entropies, we must introduce some extra bulk matter that lives outside of the observer's region. 
As a toy model, consider inserting a second probe particle from the Euclidean asymptotic boundary. We assume this probe is light enough to neglect backreaction, but heavy enough to treat in a geodesic approximation. We further assume that the light probe has $k$ different internal states, and that it is maximally entangled with some external reference system $R$ so the total state on the probe-reference system is $|\Psi^+\rangle = \frac{1}{\sqrt{k}}\sum_{j=1}^k|j\rangle_\text{probe}|j\rangle_R$. This setup, similar to the PSSY model \cite{penington_entanglement_2020}, is depicted in figure \ref{fig:renyi-jt-calc}a. 

Taking the observer's subregion to be defined as the subregion $a$ above, we now have two competing surfaces for the extremal surface homologous to $\partial a$. The first is again the empty surface, in which case only the entropy of the light probe will contribute. In particular, we have 
\begin{equation}
    Z_n = 
    \sum_b |b\, \psi_\beta(b)|^{2n} \text{ tr}(\rho_{E(a)}^{\text{bulk}}(b)^n) =\sum_b |b\, \psi_\beta(b)|^{2n} k^{1-n}
\end{equation}
in which case the saddle $b_n$ is again independent of $n$ and the entanglement spectrum is flat,
\begin{equation} \label{eq:S_null}
    S_n(a) =S_n(\rho_{E(a)}^{\text{bulk}}) = \log k, \qquad \gamma_a = \varnothing.
\end{equation}

The second candidate surface must be homologous to $\partial a$, exclude the light probe from the homology wedge $E_n(a)$, and minimize the dilaton value $2\pi\Phi$ (which plays the role of $\mathcal{A}/4G_N$ in JT gravity). This will of course depend on the light probe's position.
Noting that curves of constant $\sigma$ are geodesics in the double trumpet solution (\ref{eq:jt-solutions-trumpet}), we fix the light probe's position relative to the observer for all $\rho$ by specifying its $\sigma$-coordinate, which we denote as $\lambda \in(-\frac{1}{2}, \frac{1}{2})$.\footnote{We will choose to consider $|\lambda| < \frac{1}{2} - \frac{\ell}{2b_n}$ so that the probe sits outside of the observer's subregion in the saddle-point geometry of interest.}

\begin{figure}
    \centering
\includegraphics[width=0.9\linewidth]{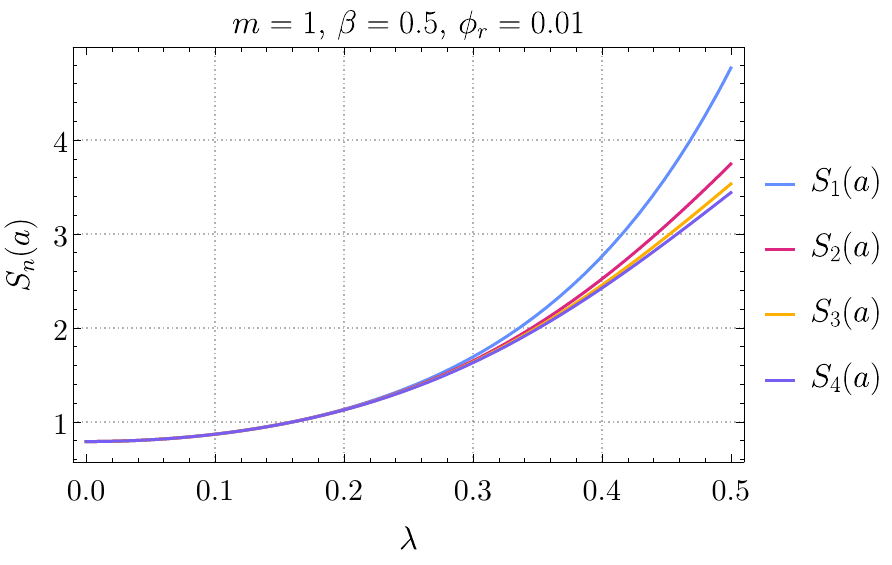}
    \caption{Rényi entropy of the observer's subregion as a function of the light probe's relative position.}
    \label{fig:renyis}
\end{figure}

\begin{figure}
    \centering
\includegraphics[width=0.9\linewidth]{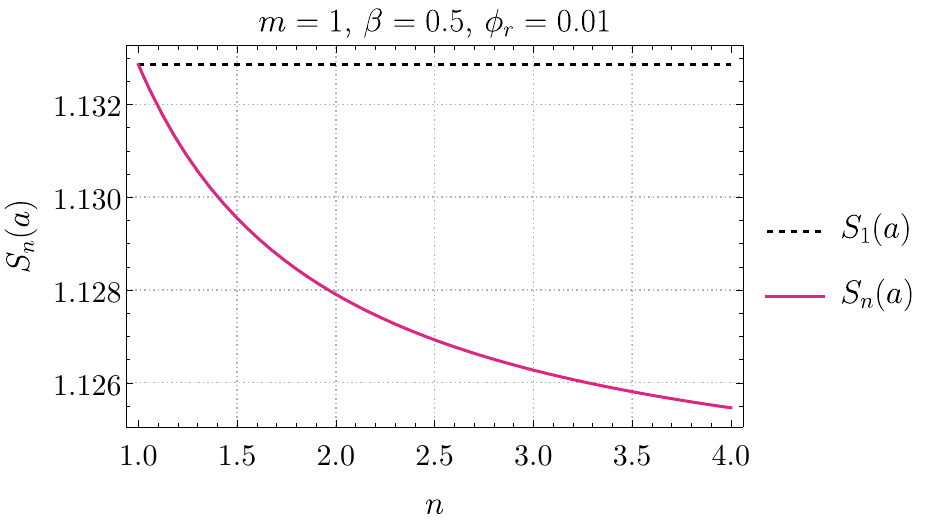}
    \caption{Behaviour of Rényi entropy as $n\to1$.}
    \label{fig:renyi-n-behaviour}
\end{figure}

Using the dilaton profile in (\ref{eq:jt-solutions-trumpet}), we find that the second candidate surface satisfying all three requirements consists of end points at $\sigma_1 = 0$ and $\sigma_2 =\lambda$ on the slice $\Sigma$ at $\rho=0$. The homology wedge $E_{n}(a)$ in this case is the portion of the slice $b$ that lies between these two values of $\sigma$; see figure \ref{fig:renyi-jt-calc}b. This has the JT area term
\begin{align}
    \frac{\mathcal{A}[b]}{4G_N} =2\pi(\Phi|_{\sigma= 0} + \Phi_0)+2\pi(\Phi|_{\sigma = \lambda} + \Phi_0) &=2S_0 +\dfrac{2\pi^2\phi_r}{\beta}\Bigg(1+\cosh(b\lambda)\Bigg) \: .
\end{align}
This area term now yields a non-trivial $b$-dependence in $Z_n$,
\begin{align}
    Z_n &= \sum_b
    |b\, \psi_\beta(b)|^{2n}\exp\Big({-(n-1)\Big[2S_0 + \frac{2\pi^2\phi_r}{\beta}(1+\cosh(b\lambda))\Big]\Big) \:} \: .
\end{align}
The saddles $b_n$ are given by extremizing the summand, found by solving
\begin{align}\label{eq:extrmum-equation-jt-calc}
    &\notag\dfrac{-m n \beta}{\cosh(\frac{b_n}{2}) \arctan(\sinh(\frac{b_n}{2}))}+ 2(n-1)\pi^2 \lambda \phi_r \sinh(b_n\lambda) + \frac{2n}{b_n} \\
    &\hspace{1.75cm}+ n\Bigg[- m\beta + 4\phi_r + 4\phi_r\sinh\Big(\frac{b_n}{2}\Big)\arctan\Big(\sinh\Big(\frac{b_n}{2}\Big)\Big)\Bigg]\tanh\Big(\frac{b_n}{2}\Big) = 0 \: .
\end{align}
With these saddles, we compute the Rényi entropy using the replica trick,
\begin{align}
    \notag S_n(a) &= \frac{n}{1-n}\log(Z_n/Z_1^n)\\
    &=\frac{2n}{n-1}\log\Bigg(\frac{b_1|\psi_\beta(b_1)|}{b_n|\psi_\beta(b_n)|}\Bigg) + \frac{2\pi^2\phi_r}{\beta}\Big(1+\cosh(b_n \lambda)\Big) + 2S_0 \: .
\end{align}
We depict the behavior of $S_n(a)$ numerically in figure \ref{fig:renyis}. The smooth convergence of the analytically continued Rényi entropy to the von Neumann entropy in the $n\to 1$ limit is also numerically illustrated in figure \ref{fig:renyi-n-behaviour}. However, the $n=1$ saddle can be found in the large mass limit,
\begin{equation} \label{eq:b1}
    b_1 \approx 2\log\left(\frac{m\beta}{\pi\phi_r}\right) \qquad (m\gg1) \: .
\end{equation}
In this large mass limit, the von Neumann entropy is
\begin{equation} \label{eq:Sa}
    S(a) = 2S_0+\frac{2\pi^2\phi_r}{\beta}\Bigg(1+\cosh\Bigg(2\lambda\log\frac{m\beta}{\pi\phi_r}\Bigg)\Bigg) \qquad (m\gg1 \: , \: \gamma_a \neq \varnothing) \: .
\end{equation}
We will denote 
\begin{equation}
    S_{\lambda}(\beta, \phi_{r}, m) \equiv \frac{2\pi^2\phi_r}{\beta} \left[ 1+\cosh\left( 2\lambda\log\frac{m\beta}{\pi\phi_r}\right)\right]
\end{equation}
in the following for convenience.

The generalized entropy of the observer occupying $a\subset\Sigma$ is then given by the minimum of either (\ref{eq:S_null}) for $\gamma_a = \varnothing$ or (\ref{eq:Sa}) for $\gamma_a\neq\varnothing$. In order for the non-trivial homologous surface to dominate, the probe's number of flavor degrees of freedom must be lower bounded by the surface's area:
\begin{equation} \label{eq:bound_k}
    \log k > 2S_0 + S_{\lambda}(\beta, \phi_{r}, m) \: .
\end{equation}
This requires the probe to have very large entropy, at least $\mathcal{O}(2S_0)$, which is likely untenable with our assumption that the mass of the probe is light enough to be non-backreacting. A more realistic model of an observer (with mass $m_1$) and light probe (with mass $m_2 \ll m_1$) was studied in \cite{usatyuk_closed_2024}, where the backreacted dilaton profile takes the form of piecewise $\cosh(\cdot)$ functions; see appendix A.3 of \cite{usatyuk_closed_2024} for more details. In this case, the generalized entropy of the observer will take a similar form with some $S_\lambda$ now dependent on $\beta$, $\phi_r$, $m_1$, and $m_2$.

We note that the size $\ell$ of the observer's subregion does not affect the entropy in this simple model where we only insert one light probe outside of this region. Instead, the entropy is entirely determined by the probe's relative position to the observer, given by the coordinate $\lambda$. Since the observer particle sits at the maximum value of the dilaton profile (\ref{eq:jt-solutions-trumpet}), the generalized entropy of the observer decreases as the light probe is inserted further away from the observer (i.e.\ for smaller values of $\lambda$).

\section{Estimating the bulk Hilbert space dimension} \label{sec:bulk_HS}

In the previous section, we applied KRR's path integral rules to find the generalized entropy of a bulk observer in a closed universe. Of course, this generalized entropy can be used to lower bound the observer's Hilbert space. In this section, we will use KRR's hollowing prescription to make two additional estimates of the full bulk Hilbert space dimension.

\subsection{Entropy of $R$} \label{sec:ref_ent}

We first use an external reference entangled with bulk degrees of freedom, following appendix B of \cite{akers_observers_2025}. Consider estimating the dimension of some Hilbert space $\hs_B$; by introducing an external reference $R$ with arbitrarily large dimension, $\dim\hs_R > \dim\hs_B$, we can consider an entangled state $|\phi\rangle$ on the bipartite system $\hs_B\otimes\hs_R$. The dimension of our mystery Hilbert space can then be determined by maximizing the entropy of $R$ over choices of $|\phi\rangle$, similar to equation (B.1) in \cite{akers_observers_2025}:
\begin{equation}\label{eq:dimension-estimate}
    \log\dim\hs_B \geq \max_{|\phi\rangle\in\hs_B\otimes\hs_R} S(R)_{|\phi\rangle} \: .
\end{equation}

In the closed universe tensor network models of \cite{akers_observers_2025} reviewed in section \ref{sec:KRR=CO},
$|\phi\rangle \in \hs_\text{eff}\otimes\hs_R$ is chosen by maximally entangling each bulk input leg with $R$. Calculating $S(R)_{|\phi\rangle}$ here would reproduce the dimension of the effective Hilbert space; instead, we apply either the unmodified tensor network $V_\text{CU}$ given in (\ref{eq:before_hollow}) or the observer-modified $V_\text{ob}$ given in (\ref{eq:after_hollow}) to obtain a fundamental density matrix:
\begin{equation}
    \rho = 
    \begin{cases}
        V_\text{CU} |\phi\rangle\langle\phi| V_\text{CU}^\dagger,   &   \text{no observer} \\
        V_\text{ob} |\phi\rangle\langle\phi| V_\text{ob}^\dagger,   &   \text{observer}
    \end{cases} \: .
\end{equation}
In appendix B of \cite{akers_observers_2025}, the second R{\'e}nyi entropy of $R$ in this fundamental states was found to be
\begin{equation}
    S_2(R)_\rho \approx 
    \begin{cases}
        0,  &   \text{no observer} \\
        \log(d_{x,\text{bulk}} \, d_{x,e_1} \, d_{x,e_2}),   &   \text{observer at } x
    \end{cases}
\end{equation}
where a minimization was taken over a cut through the network homologous to the observer's boundary. Taking $d_{x,e_1} = d_{x,e_2} = e^{S_0}$ to make contact with JT gravity and relabeling $d_{x,\text{bulk}} \to d_\text{ob} $, this reproduces AAIL's result (\ref{eq:dimH_HUZ_AAIL}); identifying $d_{x,e_1} \, d_{x,e_2}$ as an area term in a one-dimensional network gives (\ref{eq:dimH_CO}).

We now apply a similar technique to the path integral using KRR's rules for the bulk observer. To entangle the bulk effective degrees of freedom on $\Sigma$ with a reference $R$, we imagine inserting light probe particles with internal degrees of freedom entangled with $R$, just as in the JT gravity toy  model used in section \ref{sec:Sgen_obs_JT}. Rather than a single such probe particle used there, we will insert a second non-backreacting probe -- one on each side of the observer -- to further maximize the entanglement with $R$.

\begin{figure}
    \centering
    \input{tikz/two-probes}
    \caption{Toy model to calculate the Hilbert space dimension of a closed JT universe using the KRR path integral rules.}
    \label{fig:multiple-particles}
\end{figure}
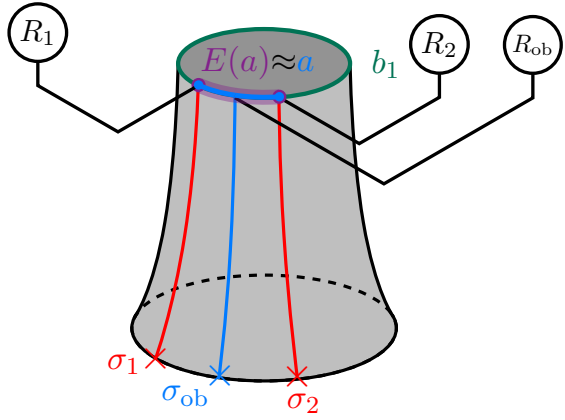

Let us thus consider the two probe particles inserted at the Euclidean asymptotic boundary at coordinates $\sigma_1,\sigma_2\in(-\frac{1}{2},\frac{1}{2})$. For simplicity, we assume that $\sigma_1$ and $\sigma_2$ are the same distance away from the observer particle at $\sigma_\text{ob}=\pm1/2$ so that $\sigma_2 = -\sigma_1$ and $\cosh(b\sigma_1)=\cosh(b\sigma_2)$. We maximally entangle both the probe particles and observer insertion with an external reference so that the total reference system is composed of all three purifying subsystems, $\mathcal{H}_R = \mathcal{H}_{R_\text{ob}} \otimes \mathcal{H}_{R_1}\otimes\mathcal{H}_{R_2}$. We will take each particle to have the same number $k$ of internal states. Our toy model is depicted in figure \ref{fig:multiple-particles}.

Following (\ref{eq:dimension-estimate}), we now apply KRR's rules to compute the generalized entropy of the reference $R$. In the fixed geometry basis, this amounts to performing the replica trick on the reduced state
\begin{equation}
    \rho_R(h) = \tr_a V_a |h\rangle \langle h| V^\dagger_a
\end{equation}
with $h$ given by the geodesic length basis $b$ above. For the $n$-th R\'enyi entropy, this requires cyclically gluing $R$ and identically gluing the observer's subregion $a$ among the $n$ copies. We also permit all possible gluings on $\Sigma\backslash a$, leading to two competing extremal surfaces. The first is $\gamma_R = \varnothing$ where $\Sigma\backslash a$ is glued identically, excluding the entirety of $\Sigma$ from the homology wedge of $R$. In this case, 
\begin{equation}
    S_n(R) = 3\log k, \qquad \gamma_a = \varnothing.
\end{equation}

In the second case, $\gamma_R \neq \varnothing$ so that some portion of $\Sigma$ outside the observer's subregion is included in the homology wedge of $R$. We want this wedge to include the two probes (but not the observer) in order to minimize the entropy of $\rho_{E(R)}^\text{bulk}(h)$. In order to also minimize the area (dilaton value) of the homologous surface, the end points will be anchored at the position of the two inserted probes, $\sigma_1$ and $\sigma_2$. This yields the JT area term
\begin{align*}
    \mathcal{A}[b]/4G_N &\approx 2\pi\Big(\Phi|_{ \sigma_1}+\Phi_0\Big) + 2\pi\Big(\Phi|_{ \sigma_2} +\Phi_0\Big)\\
    &= 2S_0+\dfrac{4\pi^2\phi_r}{\beta}\cosh(\sigma_1b) \: ,  \numberthis
\end{align*}
where in the second line we used the dilaton profile (\ref{eq:jt-solutions-trumpet}) and the symmetry assumption $\sigma_2 = -\sigma_1$.

Furthermore, the bulk entropy of the homology wedge is just that of the observer,
\begin{equation}
    \text{tr}((\rho_{E(R)}^{\text{bulk}})^n) = k^{1-n} \: .
\end{equation}
Combining the above results yields
\begin{equation}\label{eq:tr-rho-n-many-particles}
    Z_n \approx \sum_b|b\,\psi_\beta(b)|^{2n}\exp\Bigg(-(n-1)\Bigg[2S_0+\frac{ 4\pi^2\phi_r}{\beta}\cosh(b\sigma_1)\Bigg]\Bigg) k^{(1-n)} \: .
\end{equation}
One can extremize the above equation to find the saddles  $b_n$ for each $n$, in which case we get the following expression for the Rényi entropy,
\begin{equation}
    S_n(R) = \dfrac{2n}{n-1}\log\Bigg(\dfrac{|b_1\psi_\beta(b_1)|}{|b_n\psi_\beta(b_n)|}\Bigg) +2S_0 +\dfrac{4\pi^2\phi_r}{\beta}\cosh(b_n\sigma_1) + \log k \: .
\end{equation}
In the $n\to1$ limit, we reproduce the von Neumann of the reference system,
\begin{equation} \label{eq:SR}
    S(R) = 2S_0 + \frac{4\pi^2\phi_r}{\beta}\cosh(b_1\sigma_1) + \log k, \qquad \gamma_R \neq \varnothing
\end{equation}
Where again $b_1 \approx 2\log(m\beta/\phi_r\pi)$ in the large $m$ regime. 

If we assume that $k$ is large enough so that the empty surface does not dominate, the generalized entropy of $R$ is given by (\ref{eq:SR}). We can maximize this entropy by moving the probe particles to larger $|\sigma|$, closer to the observer. However, we are limited by the observer's boundary; recalling that $a$ is defined as the region anchored to the points on $\Sigma$ that are proper distance $l/2$ away from the observer, this implies that $\sigma_1 = - \sigma_{2} = (b_{1} - \ell)/2b_1$. Using (\ref{eq:dimension-estimate}), this allows us to bound the Hilbert space dimension of the closed JT universe by
\begin{equation} \label{eq:dim_final}
    \dim \mathcal{H}_\text{fun} \geq e^{2S_0 + \frac{4\pi^2\phi_r}{\beta} \cosh((b_1-l)/2)} k \: .
\end{equation}
Here, the exponential factor can be interpreted as the observer's area term $e^{\mathcal{A}_\text{ob}/4G_N}$, and the second factor as the dimension of the observer's Hilbert space in agreement with the CO result (\ref{eq:dimH_CO}). The $S_0$ dependence of this area factor also matches AAIL's $\mathcal{O}(e^{2S_0})$ result for the dimension of the Hilbert space relative to the observer in (\ref{eq:dimH_HUZ_AAIL}).

\subsection{Variation of the inner product} \label{sec:inner_prod}

Here, we provide a second estimate of the fundamental Hilbert space dimension using the variation of the inner product $\sigma^2$ defined in (\ref{eq:variation}),
\begin{equation} \label{eq:var_est}
    \dim\hs_\text{fun} \sim \frac{1}{\sigma^2} \: .
\end{equation}
This estimate has been used in a number of recent works, including for example \cite{harlow_quantum_2026,abdalla_gravitational_2025,akers_observers_2025}.\footnote{AAIL additionally provided a more accurate calculation of the Hilbert space dimension using the resolvent in \cite{abdalla_gravitational_2025}. While the $1/\sigma^2$ estimate used here is less precise, it is sufficient to reproduce the $S_0$ scaling of AAIL's result.}
Famously, the unmodified path integral calculation of the inner product squared in a closed universe is dominated by three terms:
\begin{equation} \label{eq:var2_unmodified}
    \overline{|\langle\phi|\psi\rangle|^2} = 
    \vcenter{\hbox{
    \input{tikz/rigatoni}
    }} + 
    \vcenter{\hbox{
    \input{tikz/penne}
    }} + 
    \vcenter{\hbox{
    \input{tikz/macaroni}
    }} + \mathcal{O}(e^{-2S_0})
\end{equation}
In JT gravity, these terms -- dubbed ``rigatoni'', ``penne'', and ``macaroni'' by \cite{akers_observers_2025} -- all contribute at $\mathcal{O}(1)$ in $S_0$. Since the penne term exactly matches $\Big|\overline{\langle\phi|\psi\rangle}\Big|^2$, it drops out of the variance, leaving the $\mathcal{O}(1)$ penne and macaroni terms. The variance is then also $\mathcal{O}(1)$, and by (\ref{eq:var_est}) we obtain a signature of the trivial Hilbert space
\begin{equation}
    \dim\hs_\text{fun} \sim \mathcal{O}(1) \: .
\end{equation}
The HUZ and AAIL path integral rules achieve a non-trivial Hilbert space dimension by modifying or removing the problematic penne and macaroni terms.

Let us now apply KRR's path integral rules to the variance of the inner product with an observer occupying a subregion $a\subset\Sigma$. Following equation (\ref{eq:hollowed_obs}), we expand the asymptotic state $|\psi\rangle$ in a basis of ``hollowed'' fixed geometry states $|h_\text{ob}\rangle$:
\begin{equation}
    V_\text{ob} |\psi\rangle = \sum_h \psi(h_\text{ob})|h_\text{ob}\rangle \: .
\end{equation}
The first moment of the inner product then becomes\footnote{
In the following, we assume that the dominant bulk geometry contributing to $\langle\phi|\psi\rangle^n$ is time-reflection symmetric so that gluing along cuts at the time-reflection symmetric slice $\Sigma$ works as above. This amounts to assuming e.g.\ that we have the same asymptotic geometry (determined by $\beta$ and $\phi_r$) but the boundary conditions may differ in their asymptotic matter insertions which introduce probe matter in the bulk.}
\begin{equation}
    \langle\phi|V_\text{ob}^\dagger V_\text{ob} |\psi\rangle = \sum_{h,h'} \phi^*(h'_\text{ob}) \psi(h_\text{ob}) \langle h'_\text{ob}|h_\text{ob} \rangle = \sum_h
    \vcenter{\hbox{
    \input{tikz/closed_ip_first}
    }} \: . 
\end{equation}
Because of the hollowing, the observer subregion (colored blue) serves as a new boundary condition that must be satisfied in the gluing along $\Sigma$. When computing entropies in section \ref{sec:KRR_CU}, this required enforcing a cyclic permutation on the observer's subregion; since we are computing inner products here, the observer should be glued identically (blue glues to blue) while the remainder $\Sigma\backslash \text{ob}$ may be glued in all possible ways. In this computation of the first moment, the dominant contribution is of course an identical gluing on all of $\Sigma$, and the inner product remains unchanged:
\begin{equation}
    \langle\phi|V_\text{ob}^\dagger V_\text{ob} |\psi\rangle = \langle\phi|\psi\rangle.
\end{equation}

Let us now consider applying KRR's rules to the second moment:
\begin{align*} \label{eq:closed_ip2_Va}
    |\langle\phi|V_\text{ob}^\dagger V_\text{ob}|\psi\rangle|^2 &= \sum_{h_1, h_2, h_3, h_4} \phi(h_{1,\text{ob}})^* \psi(h_{2,\text{ob}}) \psi(h_{3,\text{ob}})^* \phi(h_{4,\text{ob}}) \langle h_{1,\text{ob}}| h_{2,\text{ob}}\rangle \langle h_{3,\text{ob}} | h_{4,\text{ob}} \rangle \\
    &= \sum_{h_1, h_2, h_3, h_4} 
    \vcenter{\hbox{
    \input{tikz/closed_ip_second}
    }} \: . \numberthis
\end{align*}
Here we apply the gluing rules to the computation of the inner product $\langle h_{1,\text{ob}}| h_{2,\text{ob}}\rangle \langle h_{3,\text{ob}} | h_{4,\text{ob}} \rangle$. Boundary conditions are again specified by the observer's subregion (blue is glued vertically) while the remaining portions of $\Sigma\backslash \text{ob}$ are allowed to glue in all possible ways. Three choices of gluing will be dominant:\footnote{For the latter two gluings to correspond to genuine subdominant saddles, it is of course necessary that the area functional for the conical defect should have extrema other than the empty surface. This should occur generically in the presence of backreacting matter, and for small subregions in higher dimensional gravity. }
\allowdisplaybreaks
\begin{align} 
    \vcenter{\hbox{
    \input{tikz/rigatoni_mod}
    }} \qquad & \qquad 
    \begin{cases}
        h_1 = h_2 \\
        h_3 = h_4
    \end{cases} \label{eq:rigatoni}
    \\[0.5cm]
    \vcenter{\hbox{
    \input{tikz/penne_mod}
    }} \qquad & \qquad 
    \begin{cases}
    h_1|_\text{ob} = h_2|_\text{ob} \\
    h_3|_\text{ob} = h_4|_\text{ob} \\
    h_1|_{\Sigma\backslash\text{ob}} = h_4|_{\Sigma\backslash\text{ob}} \\
    h_2|_{\Sigma\backslash\text{ob}} = h_3|_{\Sigma\backslash\text{ob}}
    \end{cases} \label{eq:penne}
    \\[0.5cm]
    \vcenter{\hbox{
    \input{tikz/macaroni_mod}
    }} \qquad & \qquad 
    \begin{cases}
    h_1|_\text{ob} = h_2|_\text{ob} \\
    h_3|_\text{ob} = h_4|_\text{ob} \\
    h_1|_{\Sigma\backslash\text{ob}} = h_3|_{\Sigma\backslash\text{ob}} \\
    h_2|_{\Sigma\backslash\text{ob}} = h_4|_{\Sigma\backslash\text{ob}}
    \end{cases} \label{eq:macaroni}
\end{align}
where the brackets on the right show the conditions imposed on the sum over $h_1,\dots,h_4$ for each of the three types of gluing.\footnote{
For terms (\ref{eq:penne}) and (\ref{eq:macaroni}), we are assuming that subregions of two fixed geometry states can be glued together if the configurations of the metric and other fields are equivalent within the subregions, even if the fixed geometry states are not the same. For example, $h_1|_\text{ob} = h_2|_\text{ob}$ requires that $h_1$ and $h_2$ are equivalent within the observer's subregion, but does \textit{not} require that $h_1 = h_2$. This assumption was not required in the R\'enyi entropy calculations of section \ref{sec:GPI} due to the diagonal approximation.
}
Comparing these with (\ref{eq:var2_unmodified}), we see that (\ref{eq:rigatoni}) reproduces the rigatoni contribution to $|\langle\phi|\psi\rangle|^2$. The equivalence is exact -- consider inserting a basis of fixed geometry states $\sum_h |h\rangle\langle h|$ for each cylinder in the rigatoni term. Since the topology is unchanged,
\begin{equation}
    \chi(\text{rigatoni}) = \chi(\text{cylinder})^2 = 0
\end{equation}
this term contributes at $\mathcal{O}(1)$ to (\ref{eq:closed_ip2_Va}).

By shrinking the observer's subregion until they are removed ($\text{ob}\to\varnothing$), terms (\ref{eq:penne}) and (\ref{eq:macaroni}) reproduce the penne and macaroni terms exactly. Therefore, we interpret these as modifications to the penne and macaroni terms. In the presence of the observer ($\text{ob}\neq\varnothing$) the topology of these terms is modified dramatically from two cylinders to a single manifold with four boundaries:
\begin{equation}
    \chi(\text{modified penne}) = \chi(\text{modified macaroni}) = \chi(\text{four-boundary}) = -2 \: .
\end{equation}
and are therefore suppressed by $\mathcal{O}(e^{-2S_0})$ in (\ref{eq:closed_ip2_Va}). We note that these terms have the same topology as the higher order term that survives in the AAIL rules; see figure 1 of \cite{abdalla_gravitational_2025}. 

Since the rigatoni contribution drops out in the subtraction between $\overline{|\langle\phi|\psi\rangle|^2}$ and $\Big|\overline{\langle\phi|\psi\rangle}\Big|^2$, only the modified penne and macaroni terms contribute to $\sigma^2$ at leading order. Continuing our order of magnitude estimate, we find:
\begin{equation} \label{eq:H_bulk}
    \sigma^2 \sim \mathcal{O}(e^{-2S_0}) \qquad \implies \qquad \dim\hs_\text{fun} \sim \mathcal{O}(e^{2S_0}) \: .
\end{equation}
This matches the $\mathcal{O}(e^{2S_0})$ scaling in (\ref{eq:dim_final}) found using the entropy of an entangling reference, as well as AAIL's estimate (\ref{eq:dimH_HUZ_AAIL}). In higher dimensions, we expect the $\mathcal{O}(e^{-2S_0})$ suppression to generalize to an area term arising from a conical defect at the observer's boundary $\partial a$, leading to a Hilbert space with dimension of $\mathcal{O}(e^{\mathcal{A}_\text{ob}/4G_N})$. In this way, the KRR path integral rules realize the dependence on the area of the observer predicted by the CO tensor network rules in (\ref{eq:dimH_CO}).

\section{Generalized entanglement wedges from the AAIL rules} \label{sec:BP_point}

We have seen that the AAIL rules can be generalized by applying the KRR path integral prescription to the case of a closed universe. In this section, we will follow the logic in the other direction, and use the AAIL rules directly to obtain the generalized entanglement wedge associated with a pointlike observer in JT gravity. 

For concreteness, we will consider the JT gravity version of a partially entangled thermal state (PETS) \cite{Goel:2018ubv}, an entangled state of a two-sided Hilbert space with a heavy operator $O$ of dimension $\Delta$, interpreted as an AAIL observer, inserted in the Euclidean past. We can write this as
\begin{equation}
    | \beta_{L}, \beta_{R}, \Delta \rangle = \sum_{m, n} e^{- \beta_{L} E_{L} / 2} e^{- \beta_{R} E_{R} / 2} O_{mn} |m \rangle_L | n \rangle_R \: ,
\end{equation}
with $O_{mn}$ the matrix elements of the inserted operator $O$. Introducing the dimensionless parameter $C = \frac{\phi_{r}}{8 \pi G_{N}}$, with $\phi_{r} = \epsilon \Phi_{b}$ the renormalized boundary value of the dilaton, we will be interested in a semi-classical limit $C \gg 1$ where the gravitational path integral localizes on a single geometry (in addition to the limit $S_{0} \gg 1$ where the topological expansion truncates). 

The assumption that $O$ is a heavy operator means that $\Delta \sim C$. 
Following \cite{abdalla_gravitational_2025}, we will also assume $\Delta \gg S_{0}$ in order to ensure that we can neglect geometries with intersecting particle worldlines at lowest order in the topological expansion. 
In \cite{Goel:2018ubv}, it was observed that whenever $\beta_{L} \neq \beta_{R}$, there exists some $\Delta_{*}(\beta_{L}, \beta_{R}) \sim C$ such that the Euclidean geometry dual to the norm of the PETS has a single horizon (or local minimum for the dilaton) when $\Delta < \Delta_{*}$, and two horizons when $\Delta > \Delta_{*}$. We will restrict our attention to the case $\Delta > \Delta_{*}$. 

\begin{figure}
    \centering
    \includegraphics[width=0.4\linewidth]{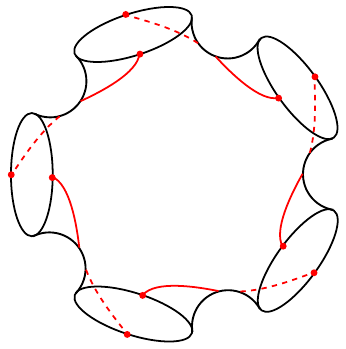}
    \caption{Pinwheel diagram for $n=5$. Red lines denote matter worldlines produced by operator insertions of dimension $\Delta$.}
    \label{fig:pinwheel}
\end{figure}

We would like to compute a gravitational entropy for the observer associated with the heavy matter particle. To begin, we should calculate the object $\text{tr}(\rho_{\text{ob}}^{n})$, where $\rho_{\text{ob}}$ is a formal object obtained from $\frac{| \beta_{L}, \beta_{R}, \Delta \rangle \langle \beta_{L}, \beta_{R}, \Delta |}{\langle \beta_{L}, \beta_{R}, \Delta | \beta_{L}, \beta_{R}, \Delta \rangle}$ by tracing out both the left and right Hilbert space. Ordinarily, this would of course just give one, but the AAIL rule tells us that the matter worldlines introduced by the operator insertions must contract cyclically between replicas. One therefore obtains
\begin{equation}
    \overline{\text{tr}(\rho_{\text{obs}}^{n})} = \frac{Z_{n}}{Z_{1}^{n}} + O(1) \: ,
\end{equation}
where $Z_{n}$ is the ``pinwheel diagram'' shown in figure \ref{fig:pinwheel}, a connected wormhole configuration with $n$ boundaries and cyclically contracted matter worldlines running between these boundaries. Subleading terms necessarily involve higher topologies, and we will only work at leading order in $e^{-2 S_{0}}$ in the following. 

The pinwheel diagram in JT gravity has been computed previously (see e.g. \cite{Boruch:2024kvv}), and is equal in our conventions to
\begin{equation}
    Z_{n} = e^{(2 - n) S_{0}} \int d k_{L} d k_{R} \: \rho_{0}(k_{L}) \rho_{0}(k_{R}) e^{- \frac{n \beta_{L} k_{L}^{2}}{2C} - \frac{n \beta_{R} k_{R}^{2}}{2C}} \gamma_{\Delta}(k_{L}, k_{R})^{n} \: ,
\end{equation}
where $\rho(k) = e^{S_{0}} \rho_{0}(k)$ is the density of states (in variable $k = \sqrt{2CE}$) 
\begin{equation}
    \rho_{0}(k) = \frac{k}{2 \pi^{2}} \sinh(2 \pi k) \: ,
\end{equation}
and
\begin{equation}
    \gamma_{\Delta}(k_{L}, k_{R}) = \frac{\prod_{\pm , \pm} \Gamma(\Delta \pm i k_{L} \pm i k_{R} )}{2^{2 \Delta - 1} C^{2 \Delta} \Gamma(2 \Delta)} \: .
\end{equation}

Suppose that the $Z_{n}$ integrals can be computed by a saddle-point approximation, with a single saddle $k_{L}^{(n)}, k_{R}^{(n)}$ dominating. The equations for these saddles are
\begin{equation}
    \begin{split}
        \partial_{k_{L}^{(n)}} \ln \rho_{0}(k_{L}^{(n)}) & = \frac{n \beta_{L} k_{L}}{C} - n \partial_{k_{L}^{(n)}} \ln \gamma_{\Delta} (k_{L}^{(n)}, k_{R}^{(n)}) \: , \\
        \partial_{k_{R}^{(n)}} \ln \rho_{0}(k_{R}^{(n)}) & = \frac{n \beta_{R} k_{R}}{C} - n \partial_{k_{R}^{(n)}} \ln \gamma_{\Delta} (k_{L}^{(n)}, k_{R}^{(n)}) \: .
    \end{split}
\end{equation}
Then we obtain annealed R{\'e}nyi entropy
\begin{multline}
    \frac{1}{1-n} \ln \left[ \overline{\text{tr}(\rho_{\text{obs}}^{n})} \right]
    = 2 S_{0} + \frac{1}{1-n} \ln \rho_{0}(k_{L}^{(n)}) - \frac{n}{1-n} \ln \rho_{0}(k_{L}^{(1)}) \\
    + \frac{1}{1-n} \ln \rho_{0}(k_{R}^{(n)}) - \frac{n}{1-n} \ln \rho_{0}(k_{R}^{(1)})
    - \frac{n \beta_{L}}{2C} \left( (k_{L}^{(n)})^{2} - (k_{L}^{(1)})^{2} \right) \\
    - \frac{n \beta_{R}}{2C} \left( (k_{R}^{(n)})^{2} - (k_{R}^{(1)})^{2} \right) - n \ln \left( \frac{\gamma_{\Delta}(k_{L}^{(n)}, k_{R}^{(n)})}{\gamma_{\Delta}(k_{L}^{(1)}, k_{R}^{(1)})} \right) \: .
\end{multline}

In the limit $n \rightarrow 1$, assuming $(k_{L}^{(n)}, k_{R}^{(n)}) \rightarrow (k_{L}^{(1)}, k_{R}^{(1)})$, we obtain (suppressing the $n=1$ superscript on $k_{L}, k_{R}$)
\begin{equation}
    S(\rho_{\text{obs}}) = 2 S_{0} + \ln \rho_{0}(k_{L}) + \ln \rho_{0}(k_{R}) \: ,
\end{equation}
where 
\begin{equation}
    \begin{split}
        \partial_{k_{L}} \ln \rho_{0}(k_{L}) & = \frac{\beta_{L} k_{L}}{C} - \partial_{k_{L}} \ln \gamma_{\Delta} (k_{L}, k_{R}) \: , \\
        \partial_{k_{R}} \ln \rho_{0}(k_{R}) & = \frac{\beta_{R} k_{R}}{C} - \partial_{k_{R}} \ln \gamma_{\Delta} (k_{L}, k_{R}) \: .
    \end{split}
\end{equation}
These are essentially thermodynamic relations, with the usual $\partial_{E} S(E) = \beta$ modified due to the coupling between left and right copies mediated by the heavy operator. 

\begin{figure}
    \centering
    \includegraphics[width=0.4\linewidth]{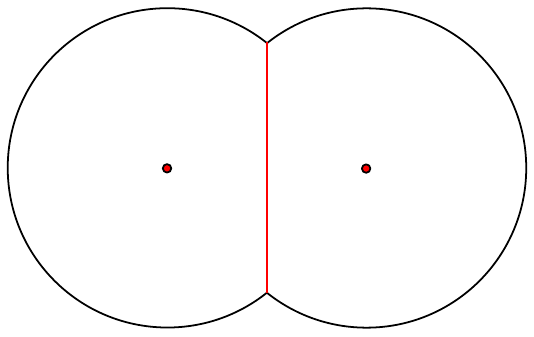}
    \caption{Bulk solution dual to the norm of the PETS state, in a case $\beta_{L} = \beta_{R}$. The particle worldline and Euclidean horizons are indicated in red.}
    \label{fig:PETS}
\end{figure}

We claim that, when the bulk solution at $n=1$ includes two Euclidean horizons separated by the observer worldline as shown in figure \ref{fig:PETS}, the quantities $\ln \rho_{0} (k_{L}), \ln \rho_{0}(k_{R})$ are approximately equal to $2 \pi$ times the values of the dilaton at the left and right horizon. This is shown in \cite{Goel:2018ubv}: the value of the dilaton at each horizon is $\Phi_{i} = \frac{2 \pi C}{\beta_{i}^{\text{eff}}}$, where $\beta_{i}^{\text{eff}}$ is the renormalized boundary length of the full disk solution from which the left or right half of the PETS geometry is a portion, and the saddle-point equations give
\begin{equation}
    k_{i} = \frac{C}{\beta_{i}^{\text{eff}}} \: , 
\end{equation}
so
\begin{equation}
    \ln \rho_{0}(k_{i}) = 2 \pi k_{i} + \mathcal{O}(\ln C ) = 2 \pi \Phi_{i} + \mathcal{O}(\ln C) \: ,
\end{equation}
as desired. We therefore obtain
\begin{equation}
    S(\rho_{\text{obs}}) = 4 \pi \Phi_{0} + 2 \pi \Phi_{L} + 2 \pi \Phi_{R} \: .
\end{equation}
This is precisely the result that would be expected for the Bousso-Penington entanglement wedge, which would include the entire interior of the black hole, with boundaries at the left and right horizon. The fact that we obtain the generalized entropy of a small neighbourhood of the pointlike observer emphasizes that, at least at the perturbative level, the AAIL rule can seemingly be viewed as a special case of the KRR prescription, since the latter can seemingly allow us to recover an entropy associated to an arbitrary bulk subregion. 

\section{Concluding remarks} \label{sec:conc}

Following the equivalence between the tensor network modifications proposed by the CO group for observers \cite{akers_observers_2025} and KRR for generalized entanglement wedges \cite{kaya_hollowgrams_2025}, we have sought to deepen the connection between related gravitational path integral modifications. We first used KRR's path integral ``hollowing'' techniques to generalize AAIL's observer rules to include extended observers occupying a bulk subregion. Conversely, we used AAIL's rules to extend KRR's derivation of the BP proposal to include generalized entanglement wedges of pointlike subregions in JT gravity.

In applying KRR's path integral rules to bulk observers, we made two estimates of the bulk Hilbert space dimension. In both (\ref{eq:dim_final}) and (\ref{eq:H_bulk}) we found non-trivial dimensions of order $\mathcal{O}(e^{2S_0})$. We emphasize that this result is not new -- both agree with the $S_0$ scaling of AAIL's result for the non-perturbative Hilbert space (\ref{eq:dimH_HUZ_AAIL}) relative to the observer. However, we find two aspects of the derivation using KRR's rules particularly interesting.

First, KRR's hollowing rules are a manifestation of AAIL's proposal that the observer's worldline serves as a conventional holographic boundary, as well as the CO assumption that observers are a part of the fundamental description. This is most clear in equation (\ref{eq:V_a}), where we interpret KRR's hollowing map $V_a$ as placing a bulk subregion $a$ on the same footing as the spatial boundary, permitting ``boundary conditions'' imposed in the path integral to be placed on both the subregion and the spatial boundary.

Second, KRR's rules make the area interpretation of AAIL's $\mathcal{O}(e^{2S_0})$ result explicit in the path integral through the introduction of conical defects. In section \ref{sec:inner_prod}, we found that these conical defects led to a suppression of the problematic penne and macaroni contributions in the second moment of the inner product (\ref{eq:var2_unmodified}). This differs from AAIL's prescription, which instead removes these problematic terms.

\subsection*{Ensemble averaging}

Finally, let us comment on interpretations of these path integral rules for observers. Throughout this work, we have followed the ``statistical'' \cite{abdalla_consistent_2026} or ``averaged holography'' \cite{harlow_observers_2026} approach, where a four-boundary path integral $G(J_1^*,J_2^*,J_3,J_4)$ with boundary conditions $J_i$ is interpreted as an average over the second moment of the inner product:
\begin{equation}
    \overline{\langle J_1|J_3\rangle\langle J_2|J_4\rangle} = G(J_1^*,J_2^*,J_3,J_4) \: .
\end{equation}
We have interpreted the KRR hollowing procedure as a modification to this second moment
\begin{equation} \label{eq:modification}
    \overline{\langle J_1|J_3\rangle\langle J_2|J_4\rangle} \to \overline{\langle J_1|V_a^\dagger V_a| J_3 \rangle \langle J_2|V_a^\dagger V_a| J_4 \rangle}
\end{equation}
with the right hand side given by equation (\ref{eq:closed_ip2_Va}). 

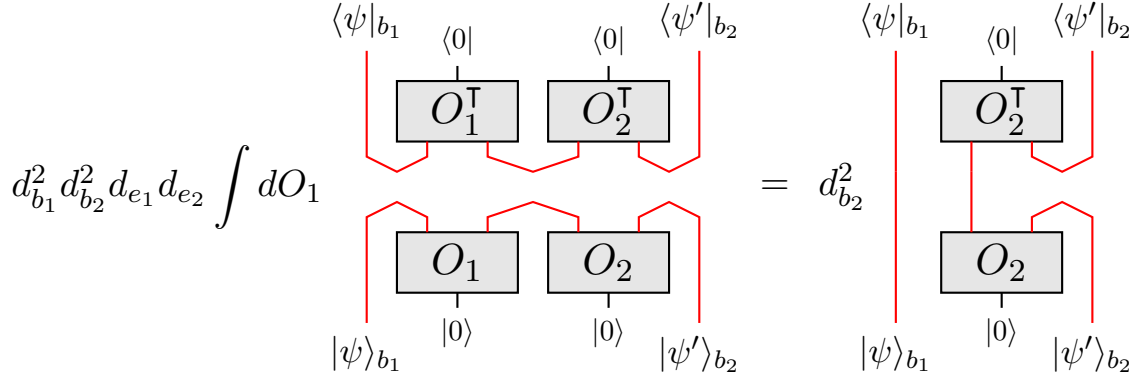
\begin{figure}
    \centering
    \input{tikz/pre-average}
    \caption{A depiction of the ``pre-averaging'' over $O_1$ that can be used to implement the removal of tensor $T_1$ acting on an observer located at bulk leg $b_1$ as prescribed by the CO rules.}
    \label{fig:pre-average}
\end{figure}

How should we interpret the right hand side of (\ref{eq:modification}) in the statistical approach to the path integral? One possibility is to achieve the modifications shown in equations (\ref{eq:rigatoni})--(\ref{eq:macaroni}) by first doing a \textit{partial} path integral over the subregion $a\subset\Sigma$ within each copy of inner product, followed by a path integral over the remainder $\Sigma\backslash a$ on the entire second moment:\footnote{An explicit definition of $G_a$ will likely make the most sense for non-asymptotic states; one possibility is to use fixed geometry states to specify the boundary conditions $J_i$. Such a definition would be analogous to the tensor network ``pre-averaging'' in equation (\ref{eq:pre-average}) and depicted in figure \ref{fig:pre-average}.}
\begin{equation} \label{eq:avg_a}
    \overline{\langle J_1|V_a^\dagger V_a| J_3 \rangle \langle J_2|V_a^\dagger V_a| J_4 \rangle} \stackrel{?}{=} G_{\Sigma\backslash a}\Big( G_a(J_1^*,J_3),\, G_a(J_2^*,J_4) \Big)
\end{equation}
This would force $a$ to connect identically between corresponding bra and ket while permitting $\Sigma\backslash a$ to connect in all possible ways as shown in equations (\ref{eq:rigatoni})--(\ref{eq:macaroni}). In this way, $G_a$ could be interpreted as a kind of ``pre-average'' over $a$ before the full path integral $G_{\Sigma\backslash a}$ is performed.\footnote{In a similar way, \cite{harlow_observers_2026} also realizes the HUZ observer rules by averaging over the boundary conditions $J$.}

A similar kind of ``pre-averaging'' can be used to understand the tensor network rules of section \ref{sec:KRR=CO}. Consider the second moment of the inner product computed using the CO tensor network rules in (\ref{eq:inner2_obsTN}). Instead of removing the tensor $T_1$ acting on the observer's bulk leg $b_1$, we could have instead averaged over the random orthogonal $O_1$ acting on $b_1$ within each copy of the inner product:
\begin{equation} \label{eq:pre-average}
    \overline{|\langle\phi|_{b_1}\langle\phi'|_{b_2}V_\text{ob}^\dagger V_\text{ob} |\psi\rangle_{b_1} |\psi'\rangle_{b_2}|^2} = \int dO_2\, \left| \left( \int dO_1 \langle\phi|_{b_1}\langle\phi'|_{b_2}V_\text{CU}^\dagger V_\text{CU} |\psi\rangle_{b_1} |\psi'\rangle_{b_2} \right) \right|^2
\end{equation}
This ``pre-averaging'' of $O_1$ -- depicted in figure \ref{fig:pre-average} -- enforces the same connection between bra and ket for the observer degrees of freedom that is accomplished by removing $T_1$ altogether. This is mathematically equivalent to averaging over the observer's operator insertion, as done by AAIL in section 5 of \cite{abdalla_gravitational_2025} and similarly by \cite{antonini_baby_2025} in the context of the Antonini-Sasieta-Swingle construction \cite{antonini_cosmology_2023}.\footnote{We thank Stefano Antonini for discussions on this subject. Upcoming work by Abdalla, Antonini, Iliesiu, and Levine will further explore this connection between observer rules and averaging \cite{abdalla_toappear}.}

\section*{Acknowledgments}

We are grateful to Chris Akers, Stefano Antonini, Sami Kaya, Alex May, Takato Mori, Pratik Rath, Martin Sasieta, Tomonori Ugajin, and Alejandro Vilar L{\'o}pez for helpful discussions and comments on earlier drafts. Research at Perimeter Institute is supported in part by the Government of Canada through the Department of Innovation, Science and  Economic Development and by the Province of Ontario through the Ministry of Colleges and Universities.

\appendix

\section{Fixed geometry states in JT gravity} \label{app:fixed_geom}

In \cite{kaya_hollowgrams_2025}, KRR note that fixed dilaton states are the appropriate incarnation of fixed geometry states in JT gravity. Here, we will review why that is the case for the two-sided JT Hilbert space, and explain why the geodesic length basis is the appropriate choice in the closed universe.

Recalling that the value of the dilaton plays the role of the area for entropy calculation in JT gravity, the fixed geometry basis must have a dilaton profile which is fixed everywhere on $\Sigma$ for every state in the basis. This can be understood to be a consequence of the gravitational constraints for a suitably chosen basis. 
To see this, let us begin by considering the two-sided JT Hilbert space. 
Let $\Sigma$ be the spacelike slice on which our state is defined, with $n^{\mu}$ a future-directed unit normal and $s^{\mu}$ a unit tangent vector to $\Sigma$. Define $\rho$ to be a proper distance coordinate on the slice, so that $s^{\mu} \nabla_{\mu} = \frac{d}{d \rho}$. We observe that the extrinsic curvature for $\Sigma$ is a number $K = s^{\mu} s^{\nu} \nabla_{\mu} n_{\nu}$, and therefore
\begin{equation}
    \begin{split}
        s_{\nu}(s^{\mu} \nabla_{\mu} s^{\nu}) & = 0 = s_{\nu} (K n^{\nu}) \\
        n_{\nu}(s^{\mu} \nabla_{\mu} s^{\nu}) & = - K = n_{\nu} (K n^{\nu}) \: ,
    \end{split}
\end{equation}
from which we infer that $s^{\mu} \nabla_{\mu} s^{\nu} = K n^{\nu}$. Similarly, $s^{\mu} \nabla_{\mu} n^{\nu} = K s^{\nu}$.

Recall that the dilaton variation and metric variation of the JT gravity action give
\begin{equation}
    R + 2 = 0 \: , \qquad \nabla_{\mu} \nabla_{\nu} \Phi - g_{\mu \nu} \Box \Phi + g_{\mu \nu} \Phi = 0
\end{equation}
respectively. Notably, contracting the latter with $g^{\mu \nu}$ yields $\Box \Phi = 2 \Phi$, which in turn gives
\begin{equation}
    \nabla_{\mu} \nabla_{\nu} \Phi = g_{\mu \nu} \Phi \: .
\end{equation}
Now, starting from the identity
\begin{equation}
    \partial_{\rho}^{2} \Phi = s^{\mu} \nabla_{\mu} (s^{\nu} \nabla_{\nu} \Phi ) = s^{\mu} s^{\nu} \nabla_{\mu} \nabla_{\nu} \Phi + s^{\mu} (\nabla_{\mu} s^{\nu}) \nabla_{\nu} \Phi \: ,
\end{equation}
we can use the equation for the dilaton and the relation $s^{\mu} \nabla_{\mu} s^{\nu} = K n^{\nu}$ to write
\begin{equation}
    \partial_{\rho}^{2} \Phi = \Phi + K n^{\nu} \nabla_{\nu} \Phi \: . 
\end{equation}
This is the first relation we will use below. To derive the second, we can begin with the identity
\begin{equation}
    \partial_{\rho} (n^{\nu} \nabla_{\nu} \Phi) = s^{\mu} \nabla_{\mu} (n^{\nu} \nabla_{\nu} \Phi) = s^{\mu} n^{\nu} \nabla_{\mu} \nabla_{\nu} \Phi + s^{\mu} (\nabla_{\mu} n^{\nu}) \nabla_{\nu} \Phi \: ,
\end{equation}
then apply the equation for the dilaton and the relation $s^{\mu} \nabla_{\mu} n^{\nu} = K s^{\nu}$ to write
\begin{equation}
    \partial_{\rho} (n^{\nu} \nabla_{\nu} \Phi) = K s^{\nu} \nabla_{\nu} \Phi \: .
\end{equation}
This is the second relation we will use below. 

We have established the constraints
\begin{equation}
    \partial_{\rho}^{2} \Phi - K \Phi_{n} - \Phi = 0 \: , \qquad \partial_{\rho} \Phi_{n} - K \partial_{\rho} \Phi = 0 \: ,
\end{equation}
where we are denoting $\Phi_{n} = n^{\mu} \nabla_{\mu} \Phi$. 
These make clear that the initial data $h_{\rho \rho}, K, \Phi, \Phi_{n}$ on $\Sigma$ are not all independent, and different bases will correspond to different choices of which of these data to fix. The fixed dilaton states of \cite{harlow2019factorization} are defined by the choice $\Phi_{n} = 0$ and a fixed horizon value for the dilaton
\begin{equation}
    \Phi(0) = \Phi_{h} \: , \quad \Phi'(0) = 0 \: ,
\end{equation}
in which case the first constraint implies
\begin{equation}
    \Phi = \Phi_{h} \cosh(\rho)
\end{equation}
while the second says that $K = 0$ except perhaps at $\rho = 0$ (where $\partial_{\rho} \Phi$ vanishes). However, the value of $K$ at $\rho = 0$ is not fixed by these constraints. It follows that the fixed dilaton basis indeed completely specifies the dilaton profile on $\Sigma$. On the other hand, the geodesic length basis of \cite{harlow2019factorization} is defined by the choice $K = 0$, in which case the constraints become
\begin{equation}
    \partial_{\rho}^{2} \Phi = \Phi \: , \qquad \partial_{\rho} \Phi_{n} = 0 \: ,
\end{equation}
which are not sufficient to completely specify the dilaton profile.

At last, let's think about the case of a closed universe with a single matter particle. The constraints in this case are the same, except that there should be a delta function-localized source for the Hamiltonian constraint, of the form $m \delta(\rho - \rho_{\text{particle}})$. Now, if we consider the geodesic length basis defined by $K = 0$, and the state in this basis defined to have length $b$, then the first constraint gives $\partial_{\rho}^{2} \Phi = \Phi$ away from the particle and $\partial_{\rho} \Phi|_{+} - \partial_{\rho} \Phi|_{-} = m$ at the particle. If the particle is located at $\rho = \pm \frac{b}{2}$ (which are identified), this equation gives $\Phi(\rho) = A \cosh(\rho) + B \sinh(\rho)$, and the jump condition then gives 
\begin{equation}
    2 A \sinh(b/2) = m \: ,
\end{equation}
while continuity at the particle gives $B = 0$. 
Thus, we obtain
\begin{equation}
    \Phi(\rho) = \frac{m}{2 \sinh(b/2)} \cosh(\rho) \: .
\end{equation}
It follows that the dilaton profile is completely fixed on $\Sigma$ in this basis.

\bibliographystyle{JHEP}
\bibliography{refs}

\end{document}

%% file: tikz/CU_TN_2.tex
\begin{tikzpicture}[very thick,scale=1]

\node[scale=1.2, anchor=east] at (-0.5,0.5) {$d_{b_1} d_{b_2} \sqrt{d_{e_1} d_{e_2}}$};

\draw[fill=gray!20] (0,0) rectangle (2,1);
\node[scale=1.5] at (1,0.5) {$O_1$};
\draw[fill=gray!20] (2.5,0) rectangle (4.5,1);
\node[scale=1.5] at (3.5,0.5) {$O_2$};

\draw (1,-0.25) -- (1,0);
\node[scale=1, anchor=north] at (1,-0.25) {$|0\rangle$};
\draw (3.5,-0.25) -- (3.5,0);
\node[scale=1, anchor=north] at (3.5,-0.25) {$|0\rangle$};

\draw[Kred] (-0.5,-0.5) -- (-0.5,1.25) -- (0,1.5) -- (0.5,1.25) -- (0.5,1);
\node[scale=1.2, anchor=north] at (-0.5,-0.5) {$b_1'$};
\node[scale=1.2, anchor=south] at (0.5,1.25) {$b_1$};

\draw[Kred] (1.5,1) -- (1.5,1.25) -- (2.25,1.5) -- (3,1.25) -- (3,1);
\node[scale=1] at (2.25,1.7) {$\langle\text{MAX}|_{e_1e_2}$};

\draw[Kred] (4,1) -- (4,1.25) -- (4.5,1.5) -- (5,1.25) -- (5,-0.5);
\node[scale=1.2, anchor=north] at (5,-0.5) {$b_2'$};
\node[scale=1.2, anchor=south] at (4,1.25) {$b_2$};

\end{tikzpicture}

%% file: tikz/Obs_2.tex
\begin{tikzpicture}[very thick,scale=1.2]

\node[scale=1.5, anchor=east] at (-1,0.5) {$d_{b_2}$};

\draw[fill=gray!20] (0,0) rectangle (2,1);
\node[scale=1.75] at (1,0.5) {$O_2$};

\draw (1,-0.25) -- (1,0);
\node[scale=1, anchor=north] at (1,-0.25) {$|0\rangle$};

\draw[Kred] (-0.75,-0.5) -- (-0.75,1.5);
\node[scale=1.2, anchor=north] at (-0.75,-0.5) {$b_1$};

\draw[Kred] (0.5,1) -- (0.5,1.5);
\node[scale=1.2, anchor=south] at (0.5,1.5) {$e_2$};

\draw[Kred] (1.5,1) -- (1.5,1.25) -- (2,1.5) -- (2.5,1.25) -- (2.5,-0.5);
\node[scale=1.2, anchor=north] at (2.5,-0.5) {$b_2'$};
\node[scale=1.2, anchor=south] at (1.5,1.25) {$b_2$};

\end{tikzpicture}

%% file: tikz/h_inner.tex
\begin{tikzpicture}[very thick, scale=1.4]

\draw[fill=gray!50]
    plot [smooth, tension=1] coordinates {(0,0) (1,0.5) (2,0)}
    plot [smooth, tension=1] coordinates {(0,0) (1,-0.5) (2,0)};

\filldraw[color=black] (0,0) circle (0.05);
\filldraw[color=black] (2,0) circle (0.05);

\node[scale=1.1, anchor=south] at (1,0.5) {$h'$};
\node[scale=1.1, anchor=north] at (1,-0.5) {$h$};
    
\end{tikzpicture}

%% file: tikz/psi_h.tex
\begin{tikzpicture}[very thick, scale=1.4]

\fill[fill=gray!50, line cap=round]
    (1,0) arc[start angle=0, end angle=-180, radius=1]
    plot [smooth,tension=1] coordinates {(-1,0) (-0.4,0.1) (0.4,-0.1) (1,0)};

\draw (1,0) arc[start angle=0, end angle=-180, radius=1];

\draw[Kdarkgreen, ultra thick, line cap=round] plot [smooth,tension=1] coordinates {(-1,0) (-0.4,0.1) (0.4,-0.1) (1,0)};

\node[scale=1.4,color=Kdarkgreen] at (-0.65,-0.1) {$\Sigma$};
\node[scale=1.1,color=Kdarkgreen,anchor=south] at (0,0) {$\langle h|$};
\node[scale=1.1,anchor=north] at (0,-1) {$|\psi\rangle$};

\end{tikzpicture}

%% file: tikz/psi_h_2.tex
\begin{tikzpicture}[very thick, scale=1.4]

\draw[fill=gray!50] (0,0) circle (1);

\draw[Kdarkgreen, ultra thick, line cap=round] plot [smooth,tension=1] coordinates {(-1,0) (-0.4,0.1) (0.4,-0.1) (1,0)};

\node[scale=1.4,color=Kdarkgreen] at (-0.65,-0.1) {$\Sigma$};

\draw[->,color=Kdarkgreen] plot [smooth, tension=1] coordinates {(1.25,0.5) (0.8,0.35) (0.5,0)};
\node[scale=1.1,color=Kdarkgreen, anchor=west] at (1.25,0.5) {$g|_\Sigma = h$};

\end{tikzpicture}

%% file: tikz/tr_rho2_h.tex
\begin{tikzpicture}[ultra thick, scale=1.4]

\begin{scope}[shift={(0,0.5)}]
    \draw[Kblue] (0,0) -- (1.2,0);
    \draw[Kpurple] (1.2,0) -- (2,0);
    \filldraw[Kblue] (0,0) circle (0.05);
    \filldraw[Kpurple] (2,0) circle (0.05);
    \draw[Kpurple, line cap = round] (1.2,-.05) -- (1.2,0.05);
    \node[scale=1.2, Kdarkgreen] at (1.75,0.25) {$h$};
    \node[scale=1] at (1.2,-0.25) {$\gamma_A$};
    \node[scale=1, Kpurple] at (2.25,0) {$A$};
    \node[scale=1, Kblue] at (-0.25,0) {$\bar{A}$};
\end{scope}

\begin{scope}[shift={(0,-0.5)}]
    \draw[Kblue] (0,0) -- (1.2,0);
    \draw[Kpurple] (1.2,0) -- (2,0);
    \filldraw[Kblue] (0,0) circle (0.05);
    \filldraw[Kpurple] (2,0) circle (0.05);
    \draw[Kpurple, line cap = round] (1.2,-.05) -- (1.2,0.05);
    \node[scale=1.2, Kdarkgreen] at (1.75,-0.25) {$h$};
\end{scope}

\begin{scope}[shift={(3,0.5)}]
    \draw[Kblue] (0,0) -- (1.2,0);
    \draw[Kpurple] (1.2,0) -- (2,0);
    \filldraw[Kblue] (0,0) circle (0.05);
    \filldraw[Kpurple] (2,0) circle (0.05);
    \draw[Kpurple, line cap = round] (1.2,-.05) -- (1.2,0.05);
    \node[scale=1.2, Kdarkgreen] at (1.75,0.25) {$h$};
\end{scope}

\begin{scope}[shift={(3,-0.5)}]
    \draw[Kblue] (0,0) -- (1.2,0);
    \draw[Kpurple] (1.2,0) -- (2,0);
    \filldraw[Kblue] (0,0) circle (0.05);
    \filldraw[Kpurple] (2,0) circle (0.05);
    \draw[Kpurple, line cap = round] (1.2,-.05) -- (1.2,0.05);
    \node[scale=1.2, Kdarkgreen] at (1.75,-0.25) {$h$};
\end{scope}

\draw[very thick, <->, Kblue] (0,-0.35) -- (0,0.35); 
\draw[very thick, <->, Kblue] (3,-0.35) -- (3,0.35); 

\draw[very thick, <->, Kpurple] plot [smooth] coordinates {(2,-0.35) (2.5,-0.1) (4.5,0.1) (5,0.35)};
\draw[very thick, <->, Kpurple] plot [smooth] coordinates {(2,0.35) (2.5,0.1) (4.5,-0.1) (5,-0.35)};
    
\end{tikzpicture}

%% file: tikz/h_before_after.tex
\begin{tikzpicture}[ultra thick, scale=1.4]

\draw[Kdarkgreen] (0,0) -- (2,0);
\filldraw[Kblue] (0,0) circle (0.05);
\filldraw[Kblue] (2,0) circle (0.05);

\node[scale=1.2, Kdarkgreen] at (1.75,0.25) {$h$};

\draw[thin, ->] (2.35,0) -- (2.65,0);

\draw[Kdarkgreen] (3,0) -- (5,0);
\filldraw[Kblue] (3,0) circle (0.05);
\filldraw[Kblue] (5,0) circle (0.05);
\draw[line width = 4.25, Kblue, line cap=round] (3.7,0) -- (4.3,0);
\node[scale=1.2, Kblue] at (4,-0.25) {$a$};

\node[scale=1.2, Kdarkgreen] at (4.75,0.25) {$h$};
    
\end{tikzpicture}

%% file: tikz/tr_rhoa2.tex
\begin{tikzpicture}[ultra thick, scale=1.4]

\begin{scope}[shift={(0,0.5)}]
    \draw[Kblue] (0,0) -- (2,0);
    \draw[Kpurple] (0.3,0) -- (1.7,0);
    \draw[Kpurple, line width = 4.25, line cap = round] (0.7,0) -- (1.3,0);
    \filldraw[Kblue] (0,0) circle (0.05);
    \filldraw[Kblue] (2,0) circle (0.05);
    \draw[Kpurple, line cap = round] (0.3,-.05) -- (0.3,0.05);
    \draw[Kpurple, line cap = round] (1.7,-.05) -- (1.7,0.05);
    \node[scale=1.2, Kdarkgreen] at (1.9,0.25) {$h$};
    \node[scale=1.2, Kpurple] at (1,0.25) {$a$};
    \node[scale=1] at (0.3,-0.25) {$\gamma_a$};
    \node[scale=1] at (1.7,-0.25) {$\gamma_a$};
\end{scope}

\begin{scope}[shift={(0,-0.5)}]
    \draw[Kblue] (0,0) -- (2,0);
    \draw[Kpurple] (0.3,0) -- (1.7,0);
    \draw[Kpurple, line width = 4.25, line cap = round] (0.7,0) -- (1.3,0);
    \filldraw[Kblue] (0,0) circle (0.05);
    \filldraw[Kblue] (2,0) circle (0.05);
    \draw[Kpurple, line cap = round] (0.3,-.05) -- (0.3,0.05);
    \draw[Kpurple, line cap = round] (1.7,-.05) -- (1.7,0.05);
    \node[scale=1.2, Kdarkgreen] at (1.9,-0.25) {$h$};
\end{scope}

\begin{scope}[shift={(3,0.5)}]
    \draw[Kblue] (0,0) -- (2,0);
    \draw[Kpurple] (0.3,0) -- (1.7,0);
    \draw[Kpurple, line width = 4.25, line cap = round] (0.7,0) -- (1.3,0);
    \filldraw[Kblue] (0,0) circle (0.05);
    \filldraw[Kblue] (2,0) circle (0.05);
    \draw[Kpurple, line cap = round] (0.3,-.05) -- (0.3,0.05);
    \draw[Kpurple, line cap = round] (1.7,-.05) -- (1.7,0.05);
    \node[scale=1.2, Kdarkgreen] at (1.9,0.25) {$h$};
\end{scope}

\begin{scope}[shift={(3,-0.5)}]
    \draw[Kblue] (0,0) -- (2,0);
    \draw[Kpurple] (0.3,0) -- (1.7,0);
    \draw[Kpurple, line width = 4.25, line cap = round] (0.7,0) -- (1.3,0);
    \filldraw[Kblue] (0,0) circle (0.05);
    \filldraw[Kblue] (2,0) circle (0.05);
    \draw[Kpurple, line cap = round] (0.3,-.05) -- (0.3,0.05);
    \draw[Kpurple, line cap = round] (1.7,-.05) -- (1.7,0.05);
    \node[scale=1.2, Kdarkgreen] at (1.9,-0.25) {$h$};
\end{scope}

\draw[very thick, <->, Kblue] (0,-0.35) -- (0,0.35); 
\draw[very thick, <->, Kblue] (2,-0.35) -- (2,0.35); 
\draw[very thick, <->, Kblue] (3,-0.35) -- (3,0.35); 
\draw[very thick, <->, Kblue] (5,-0.35) -- (5,0.35); 

\draw[very thick, <->, Kpurple] plot [smooth] coordinates {(1,-0.35) (1.5,-0.1) (3.5,0.1) (4,0.35)};
\draw[very thick, <->, Kpurple] plot [smooth] coordinates {(1,0.35) (1.5,0.1) (3.5,-0.1) (4,-0.35)};
    
\end{tikzpicture}

%% file: tikz/closed_state.tex
\begin{tikzpicture}[very thick, scale=1.4]

\fill[gray!50]
    (0,2) ellipse (0.65 and 0.26)
    (0,0) ellipse (1 and 0.4);

\fill[gray!50]
    plot [smooth,tension=1] coordinates {(0.65,2) (0.75,0.8) (0.977,0.085)} -- plot [smooth,tension=1] coordinates {(-0.977,0.085) (-0.75,0.8) (-0.65,2)} -- cycle;

\begin{scope}
\clip (-1.1,0.1) rectangle (1.1,-0.5);
\draw (0,0) ellipse (1 and 0.4);
\end{scope}
\draw[dashed] (0,0) ellipse (1 and 0.4);

\draw[line cap = round] 
    plot [smooth,tension=1] coordinates {(0.65,2) (0.75,0.8) (0.977,0.085)}
    plot [smooth,tension=1] coordinates {(-0.65,2) (-0.75,0.8) (-0.977,0.085)};

\draw[Kdarkgreen, ultra thick, fill=gray!85] (0,2) ellipse (0.65 and 0.26);

\node[scale=1.2, anchor=north] at (0,-0.4) {$|\psi\rangle$};
\node[Kdarkgreen, scale=1.2, anchor=south] at (0,2.26) {$\langle h|$};
\node[Kdarkgreen, scale=1.2, anchor=east] at (-0.65,2) {$\Sigma$};
    
\end{tikzpicture}

%% file: tikz/closed_Vobs.tex
\begin{tikzpicture}[very thick, scale=1.4]

\fill[gray!50]
    (0,2) ellipse (0.65 and 0.26)
    (0,0) ellipse (1 and 0.4);

\fill[gray!50]
    plot [smooth,tension=1] coordinates {(0.65,2) (0.75,0.8) (0.977,0.085)} -- plot [smooth,tension=1] coordinates {(-0.977,0.085) (-0.75,0.8) (-0.65,2)} -- cycle;

\begin{scope}
\clip (-1.1,0.1) rectangle (1.1,-0.5);
\draw (0,0) ellipse (1 and 0.4);
\end{scope}
\draw[dashed] (0,0) ellipse (1 and 0.4);

\draw[line cap = round] 
    plot [smooth,tension=1] coordinates {(0.65,2) (0.75,0.8) (0.977,0.085)}
    plot [smooth,tension=1] coordinates {(-0.65,2) (-0.75,0.8) (-0.977,0.085)};

\draw[Kdarkgreen, ultra thick, fill=gray!85] (0,2) ellipse (0.65 and 0.26);

\draw[line width=4.25, Kblue, line cap = round] 
    (-0.65,2) arc [start angle=180, end angle=110, x radius=0.65, y radius = 0.26]
    (-0.65,2) arc [start angle=180, end angle=250, x radius=0.65, y radius = 0.26];

\node[scale=1.2, Kblue] at (-0.7,2.35) {ob};
\node[scale=1.2, Kdarkgreen] at (0.7,2.3) {$h$};
    
\end{tikzpicture}

%% file: tikz/tr_rho2_closed.tex
\begin{tikzpicture}[very thick, scale=1.4]

\newcommand{\obRing}[3]{
    \draw[#3, ultra thick] (#1,#2) ellipse (0.65 and 0.26);

    \draw[line width=4.25, Kpurple, line cap = round] 
        (#1-0.65,#2) arc [start angle=180, end angle=110, x radius=0.65, y radius = 0.26]
        (#1-0.65,#2) arc [start angle=180, end angle=250, x radius=0.65, y radius = 0.26];
}

\node[scale=1, anchor=east] at (-1.5,0) {$\tr \rho_\text{ob}(h)^2 =$};

\obRing{0}{0.5}{Kblue};
\obRing{2}{0.5}{Kblue};
\obRing{0}{-0.5}{Kblue};
\obRing{2}{-0.5}{Kblue};
\node[scale=1.2, Kpurple] at (-0.7,0.85) {ob};

\draw[<->, Kpurple] plot [smooth] coordinates {(-0.6,-0.3) (-0.7,-0.1) (1.4,0.3)};
\draw[<->, Kpurple] plot [smooth] coordinates {(-0.6,0.3) (-0.7,0.1) (1.4,-0.3)};

\draw[<->, Kblue] plot [smooth] coordinates {(0.6,-0.3) (0.7,0) (0.6,0.3)};
\draw[<->, Kblue] plot [smooth] coordinates {(2.6,-0.3) (2.7,0) (2.6,0.3)};

\begin{scope}[shift={(0,-2.25)}]
    \node[scale=1.2] at (-1.25,0) {$+$};

    \obRing{0}{0.5}{Kpurple};
    \obRing{2}{0.5}{Kpurple};
    \obRing{0}{-0.5}{Kpurple};
    \obRing{2}{-0.5}{Kpurple};
    \node[scale=1.2, Kpurple] at (-0.7,0.85) {ob};
    
    \draw[<->, Kpurple] plot [smooth] coordinates {(-0.6,-0.3) (-0.7,-0.1) (1.4,0.3)};
    \draw[<->, Kpurple] plot [smooth] coordinates {(-0.6,0.3) (-0.7,0.1) (1.4,-0.3)};
    
    \draw[<->, Kpurple] plot [smooth] coordinates {(2.6,-0.3) (2.7,-0.1) (0.6,0.3)};
    \draw[<->, Kpurple] plot [smooth] coordinates {(2.6,0.3) (2.7,0.1) (0.6,-0.3)};
\end{scope}

\begin{scope}[shift={(0,-4.5)}]
    \node[scale=1.2] at (-1.25,0) {$+$};

    \obRing{0}{0.5}{Korange};
    \obRing{2}{0.5}{Korange};
    \obRing{0}{-0.5}{Korange};
    \obRing{2}{-0.5}{Korange};
    \node[scale=1.2, Kpurple] at (-0.7,0.85) {ob};
    
    \draw[<->, Kpurple] plot [smooth] coordinates {(-0.6,-0.3) (-0.7,-0.1) (1.4,0.3)};
    \draw[<->, Kpurple] plot [smooth] coordinates {(-0.6,0.3) (-0.7,0.1) (1.4,-0.3)};
    
    \draw[<->, Korange] plot [smooth] coordinates {(2.6,-0.3) (2.7,-0.05) (0.6,-0.3)};
    \draw[<->, Korange] plot [smooth] coordinates {(2.6,0.3) (2.7,0.05) (0.6,0.3)};
\end{scope}

\end{tikzpicture}

%% file: tikz/doubletrumpet-jt.tex
\begin{tikzpicture}[very thick, scale=1.4]

\fill[gray!50]
    (0,1.6) ellipse (0.65 and 0.26)
    (0,0) ellipse (1 and 0.4);

\fill[gray!50]
    plot [smooth,tension=1] coordinates {(0.65,1.6) (0.75,0.64) (0.977,0.068)} -- plot [smooth,tension=1] coordinates {(-0.977,0.068) (-0.75,0.64) (-0.65,1.6)} -- cycle;

\begin{scope}
\clip (-1.1,0.1) rectangle (1.1,-0.5);
\draw (0,0) ellipse (1 and 0.4);
\end{scope}
\draw[dashed] (0,0) ellipse (1 and 0.4);

\draw[line cap = round] 
    plot [smooth,tension=1] coordinates {(0.65,1.6) (0.75,0.64) (0.977,0.068)}
    plot [smooth,tension=1] coordinates {(-0.65,1.6) (-0.75,0.64) (-0.977,0.068)};

\begin{scope}[yshift=3.2cm, yscale=-1]

\fill[gray!50]
    plot [smooth,tension=1] coordinates {(0.65,1.6) (0.75,0.64) (0.977,0.068)} -- plot [smooth,tension=1] coordinates {(-0.977,0.068) (-0.75,0.64) (-0.65,1.6)} -- cycle;

\fill[gray!85]
    (0,0) ellipse (1 and 0.42);

\begin{scope}
\clip (-1.1,0.1) rectangle (1.1,-0.5);
\draw (0,0) ellipse (1 and 0.4);
\end{scope}
\draw (0,0) ellipse (1 and 0.4);

\draw[line cap = round] 
    plot [smooth,tension=1] coordinates {(0.65,1.6) (0.75,0.64) (0.977,0.068)}
    plot [smooth,tension=1] coordinates {(-0.65,1.6) (-0.75,0.64) (-0.977,0.068)};
\end{scope}

\draw[Kdarkgreen, ultra thick, dashed] (0,1.6) ellipse (0.65 and 0.26);

\node[Kdarkgreen, scale=1.2, anchor=east] at (-0.65,1.6) {$b_*$};

\draw[->, >=stealth, thick, yshift=-0.15cm] (270:1 and 0.4) arc [start angle=270, end angle=350, x radius=1, y radius=0.4] node[midway, below right] {$\beta$};

\draw[Kblue, line cap = round] plot [smooth, tension=1] coordinates {(-0.342, -0.376) (-0.255, 0.45) (-0.22, 1.356) (-0.255, 2.47) (-0.342, 2.817)};

\path (-0.342, -0.376) node[Kblue] {$\times$} node[below left=-1pt, Kblue] {$\mathcal{O}_i$};

\path (-0.342, 2.817) node[Kblue] {$\times$} node[below left=-1pt, Kblue] {};

\end{tikzpicture}

%% file: tikz/psi_beta.tex
\begin{tikzpicture}[very thick, scale=1.4]

\fill[gray!50]
    (0,2) ellipse (0.65 and 0.26)
    (0,0) ellipse (1 and 0.4);

\fill[gray!50]
    plot [smooth,tension=1] coordinates {(0.65,2) (0.75,0.8) (0.977,0.085)} -- plot [smooth,tension=1] coordinates {(-0.977,0.085) (-0.75,0.8) (-0.65,2)} -- cycle;

\begin{scope}
\clip (-1.1,0.1) rectangle (1.1,-0.5);
\draw (0,0) ellipse (1 and 0.4);
\end{scope}
\draw[dashed] (0,0) ellipse (1 and 0.4);

\draw[line cap = round] 
    plot [smooth,tension=1] coordinates {(0.65,2) (0.75,0.8) (0.977,0.085)}
    plot [smooth,tension=1] coordinates {(-0.65,2) (-0.75,0.8) (-0.977,0.085)};

\draw[Kblue, line cap = round] plot [smooth, tension=1] coordinates {(-0.342, -0.376) (-0.255, 0.49) (-0.22, 1.72)} node[midway, left] {};

\path (-0.342, -0.376) node[Kblue] {$\times$} node[below left=-1pt, Kblue] {$\mathcal{O}_i$};

\draw[Kdarkgreen, ultra thick, fill=gray!85, samples=100, smooth] 
    plot[domain=0:360] ({(0.64 + 0.04*sin(5*\x))*cos(\x)}, {2 + (0.26 + 0.04*sin(6*\x)) * sin(\x)});

\node[Kdarkgreen, scale=1.2, anchor=east] at (-0.65,2) {$\Sigma$};

\node[scale=1.2, anchor=north] at (0,-0.6) {$|\psi_\beta\rangle$};

\draw[->, >=stealth, thick, yshift=-0.15cm] (270:1 and 0.4) arc [start angle=270, end angle=350, x radius=1, y radius=0.4] node[midway, below right] {$\beta$};

\end{tikzpicture}

%% file: tikz/psi_b_.tex
\begin{tikzpicture}[very thick, scale=1.4]

\fill[gray!50]
    (0,2) ellipse (0.65 and 0.26)
    (0,0) ellipse (1 and 0.4);

\fill[gray!50]
    plot [smooth,tension=1] coordinates {(0.65,2) (0.75,0.8) (0.977,0.085)} -- plot [smooth,tension=1] coordinates {(-0.977,0.085) (-0.75,0.8) (-0.65,2)} -- cycle;

\begin{scope}
\clip (-1.1,0.1) rectangle (1.1,-0.5);
\draw (0,0) ellipse (1 and 0.4);
\end{scope}
\draw[dashed] (0,0) ellipse (1 and 0.4);

\draw[line cap = round] 
    plot [smooth,tension=1] coordinates {(0.65,2) (0.75,0.8) (0.977,0.085)}
    plot [smooth,tension=1] coordinates {(-0.65,2) (-0.75,0.8) (-0.977,0.085)};

\draw[Kdarkgreen, ultra thick, fill=gray!85] (0,2) ellipse (0.65 and 0.26);

\node[scale=1.2, anchor=north] at (0,-0.6) {$|\psi_\beta\rangle$};

\node[Kdarkgreen, scale=1.2, anchor=south] at (0,2.26) {$\langle b|$};

\draw[Kblue, line cap = round] plot [smooth, tension=1] coordinates {(-0.342, -0.376) (-0.255, 0.49) (-0.22, 1.72)} node[midway, left] {};

\draw[->, >=stealth, thick, yshift=-0.15cm] (270:1 and 0.4) arc [start angle=270, end angle=350, x radius=1, y radius=0.4] node[midway, below right] {$\beta$};

\path (-0.342, -0.376) node[Kblue] {$\times$} node[below left=-1pt, Kblue] {$\mathcal{O}_i$};

\end{tikzpicture}

%% file: tikz/probediagram-smaller.tex
\begin{tikzpicture}[very thick, scale=1.75]

\fill[gray!50]
    (0,2) ellipse (0.65 and 0.26)
    (0,0) ellipse (1 and 0.4);

\fill[gray!50]
    plot [smooth,tension=1] coordinates {(0.65,2) (0.75,0.8) (0.977,0.085)} -- plot [smooth,tension=1] coordinates {(-0.977,0.085) (-0.75,0.8) (-0.65,2)} -- cycle;

\begin{scope}
\clip (-1.1,0.1) rectangle (1.1,-0.5);
\draw (0,0) ellipse (1 and 0.4);
\end{scope}
\draw[dashed] (0,0) ellipse (1 and 0.4);

\draw[line cap = round] 
    plot [smooth,tension=1] coordinates {(0.65,2) (0.75,0.8) (0.977,0.085)}
    plot [smooth,tension=1] coordinates {(-0.65,2) (-0.75,0.8) (-0.977,0.085)};

\draw[Kdarkgreen, ultra thick, fill=gray!85] (0,2) ellipse (0.65 and 0.26);

\draw[Kblue, line cap = round] plot [smooth, tension=1] coordinates {(-0.342, -0.376) (-0.255, 0.59) (-0.22, 1.74)} node[midway, left] {};

\path (-0.335, -0.36) node[black, scale=1.45] {$\times$} node[below left=-1pt, Kblue, scale=1.45] {$\text{ob}$};

\draw[Kred, line cap = round] plot [smooth, tension=1] coordinates {(0.8, -0.23) (0.52, 0.7) (0.45, 1.81)};

\path (0.8, -0.25) node[Kred, scale=1.45] {$\times$};

\node[Kred, scale=1.45, anchor=north] at (0.8, -0.3) {probe};

\path (0.45, 1.81) node[Kred, scale=1.45] {$\times$} node[below left=-1pt, Kred, scale=1.45]{};

\draw[Kpurple, opacity=0.4, line width=5.625pt] (-0.22, 1.76)  
    arc (250:70:0.65 and 0.26) node[circle, fill=Kpurple, inner sep=1.875pt, opacity=1, label={[text=Kpurple, opacity=1, scale=1.25, label distance= -5pt]above:{$\sigma=0$}}] {};

\node[Kpurple, scale=1.45, anchor=south west] at (-1.45, 2) {$E(a)$};

\node[Kpurple, scale=1.25, anchor=north] at (1.1, 2.2) {$\sigma = \lambda$};

\draw[thick, black, opacity=1, line width=1.25pt] (0.45, 1.81)
    -- ++(-30:0.9)         
    -- ++(30:0.9)          
    -- ++(0, 0.4)          
    node[circle, draw=black, fill=white, line width=1.25pt, inner sep=3pt, scale=1.45] {\small R};

\draw[Kpurple, opacity=0.4, line width=5.625pt] (-0.22, 1.76) 
    arc (250:310:0.65 and 0.26) node[circle, fill=Kpurple, inner sep=1.875pt, opacity=1] {};

\draw[Kblue, line width=1.875pt] (-0.22, 1.76)  
    arc (250:220:0.65 and 0.26) node[circle, fill=Kblue, inner sep=1.25pt] {};

\draw[Kblue, line width=1.875pt] (-0.22, 1.76) 
    arc (250:280:0.65 and 0.26) node[circle, fill=Kblue, inner sep=1.25pt] {};

\node[Kblue, scale=1.45, anchor=north] at (-0.02, 1.7) {$a$};

\path (-0.204, 1.78) node[black, scale=1.45] {$\times$} node[below left=-1pt, black, scale=1.45]{};

\end{tikzpicture}

%% file: tikz/dilaton_profile.tex
\begin{tikzpicture}

\begin{axis}[
    x label style={at={(1.05,0.035)}},
    xlabel = {$\sigma$},
    ylabel = {$\Phi(\rho=0,\sigma)$},
    axis y line=middle,
    axis x line=bottom,
    ymin=0.98,
    ymax=1.15,
    xmin=-0.6,
    xmax=0.6,
    ytick={0},
    yticklabels={},
    xtick={-0.5,-0.4,0.2,0.4,0.5},
    xticklabels={$-\frac{1}{2}$,,$\lambda$,,$\frac{1}{2}$},
    xmajorgrids=true,
    grid style=dashed,
    samples=100
    ]

\addplot[domain=-0.5:0, line width=5, Kpurple!40, line cap=round] {cosh(x)} node[pos=0.5, above right, Kpurple] {$E(a)$};
\addplot[domain=0.2:0.5, line width=5, Kpurple!40, line cap=round] {cosh(x)};
\addplot[domain=-0.5:-0.4, very thick, Kblue] {cosh(x)} node[pos=0.5, below left] {$a$};
\addplot[domain=0.4:0.5, very thick, Kblue] {cosh(x)} node[pos=0.5, below right] {$a$}; 
\addplot[domain=-0.4:0.4, very thick] {cosh(x)};

\addplot[only marks, mark size=2, Kblue] coordinates {(-0.4,{cosh(0.4)})};
\addplot[only marks, mark size=2, Kblue] coordinates {(0.4,{cosh(0.4)})};

\addplot[<->, thick] coordinates {(-0.5,1) (-0.4,1)} node[pos=0.5, above] {$\frac{l}{2b}$};

\addplot[only marks, mark=x, mark size=5, very thick] coordinates {(-0.5,{cosh(-0.5)}) (0.5,{cosh(0.5)})} node[pos=0,right] {observer};

\addplot[only marks, mark size=4] coordinates {(0.2,{cosh(0.2)})} node[pos=0, above left] {probe};

\end{axis}

\end{tikzpicture}

%% file: tikz/two-probes.tex
\begin{tikzpicture}[very thick, scale=1.75]

\fill[gray!50]
    (0,2) ellipse (0.65 and 0.26)
    (0,0) ellipse (1 and 0.4);

\fill[gray!50]
    plot [smooth,tension=1] coordinates {(0.65,2) (0.75,0.8) (0.977,0.085)} -- plot [smooth,tension=1] coordinates {(-0.977,0.085) (-0.75,0.8) (-0.65,2)} -- cycle;

\begin{scope}
\clip (-1.1,0.1) rectangle (1.1,-0.5);
\draw (0,0) ellipse (1 and 0.4);
\end{scope}
\draw[dashed] (0,0) ellipse (1 and 0.4);

\draw[line cap = round] 
    plot [smooth,tension=1] coordinates {(0.65,2) (0.75,0.8) (0.977,0.085)}
    plot [smooth,tension=1] coordinates {(-0.65,2) (-0.75,0.8) (-0.977,0.085)};

\draw[Kdarkgreen, ultra thick, fill=gray!85] (0,2) ellipse (0.65 and 0.26);

\draw[Kblue, line cap = round] plot [smooth, tension=1] coordinates {(-0.342, -0.376) (-0.255, 0.59) (-0.22, 1.74)} node[midway, left] {};

\path (-0.335, -0.36) node[Kblue, scale=1.45] {$\times$} node[below left=-1pt, Kblue, scale=1.15] {$\sigma_\text{ob}$};

\draw[Kred, line cap = round] plot [smooth, tension=1] coordinates {(0.25, -0.364) (0.15, 0.7) (0.113,1.744)};

\path (0.25, -0.364) node[Kred, scale=1.45] {$\times$};

\draw[Kred, line cap = round] plot [smooth, tension=1] coordinates {(-0.82, -0.23) (-0.6, 0.7) (-0.498, 1.833)};

\path (-0.82, -0.23)  node[Kred, scale=1.45] {$\times$};

\node[Kred, scale=1.15, anchor=north] at (-1.06, -0.1) {$\sigma_1$};
\node[Kred, scale=1.15, anchor=north] at (0.3, -0.4) {$\sigma_2$};

\draw[Kpurple, opacity=0.4, line width=5.625pt] (-0.22, 1.76)  
    arc (250:220:0.65 and 0.26) node[circle, fill=Kpurple, inner sep=1.875pt, opacity=1] {};

\draw[thick, black, opacity=1, line width=1.25pt] (0.113,1.744)
    -- ++(-30:0.7)         
    -- ++(30:0.7)          
    -- ++(0, 0.4)          
    node[circle, draw=black, fill=white, line width=1.25pt, inner sep=1.5pt, scale=1.15] {\small $R_2$};

\draw[thick, black, opacity=1, line width=1.25pt] (-0.22, 1.74)
    -- ++(-30:1.3)         
    -- ++(30:1.3)          
    -- ++(0, 0.4)          
    node[circle, draw=black, fill=white, line width=1.25pt, inner sep=1.5pt, scale=0.95] {\small $R_{\text{ob}}$};

\draw[thick, black, opacity=1, line width=1.25pt] (-0.498, 1.833)
    -- ++(210:0.7)         
    -- ++(150:0.7)          
    -- ++(0, 0.4)          
    node[circle, draw=black, fill=white, line width=1.25pt, inner sep=1.5pt, scale=1.15] {\small $R_1$};

\draw[Kpurple, opacity=0.4, line width=5.625pt] (-0.22, 1.76) 
    arc (250:280:0.65 and 0.26) node[circle, fill=Kpurple, inner sep=1.875pt, opacity=1] {};

\draw[Kblue, line width=1.875pt] (-0.22, 1.76)  
    arc (250:220:0.65 and 0.26) node[circle, fill=Kblue, inner sep=1.25pt] {};

\draw[Kblue, line width=1.875pt] (-0.22, 1.76) 
    arc (250:280:0.65 and 0.26) node[circle, fill=Kblue, inner sep=1.25pt] {};

\node[Kblue, scale=1.15, anchor=north] at (-0.05, 2.23) {$\textcolor{Kpurple}{E(a)} \textcolor{black}{\approx } a$};

\node[Kdarkgreen, scale=1.15, anchor=north] at (0.92, 2.2) {$b_1$};
    
\end{tikzpicture}

%% file: tikz/rigatoni.tex
\begin{tikzpicture}[very thick, scale=0.7]

\fill[fill=gray!50] 
    (1,3.5) -- (1,0) -- (1,0) arc [start angle=0, end angle=-180, x radius=1, y radius=0.4] -- (-1,3.5)
    (3.5,3.5) -- (3.5,0) -- (3.5,0) arc [start angle=0, end angle=-180, x radius=1, y radius=0.4] -- (1.5,3.5);

\draw[fill=gray!85] 
    (0,3.5) ellipse (1 and 0.4)
    (2.5,3.5) ellipse (1 and 0.4);

\draw[fill=gray!50] 
    (1,0) arc [start angle=0, end angle=-180, x radius=1, y radius=0.4]
    (3.5,0) arc [start angle=0, end angle=-180, x radius=1, y radius=0.4];

\draw[fill=gray!50, dashed] 
    (1,0) arc [start angle=0, end angle=180, x radius=1, y radius=0.4]
    (3.5,0) arc [start angle=0, end angle=180, x radius=1, y radius=0.4];

\draw
    (1,0) -- (1,3.5)
    (-1,0) -- (-1,3.5)
    (3.5,0) -- (3.5,3.5)
    (1.5,0) -- (1.5,3.5);

\node[scale=1] at (0,-0.8) {$|\psi\rangle$};
\node[scale=1] at (0,4.3) {$\langle\phi|$};
\node[scale=1] at (2.5,-0.8) {$|\phi\rangle$};
\node[scale=1] at (2.5,4.3) {$\langle\psi|$};

\end{tikzpicture}

%% file: tikz/penne.tex
\begin{tikzpicture}[very thick, scale=0.7]

\fill[fill=gray!50] 
    (1,3.5) -- (3.5,0) -- (3.5,0) arc [start angle=0, end angle=-180, x radius=1, y radius=0.4] -- (-1,3.5);

\draw
    (3.475,0.09) -- (0.975,3.59)
    (1.525,-0.09) -- (-0.975,3.41);
    
\fill[fill=gray!50]  
    (3.5,3.5) -- (1,0) -- (1,0) arc [start angle=0, end angle=-180, x radius=1, y radius=0.4] -- (1.5,3.5);

\draw[fill=gray!85] 
    (0,3.5) ellipse (1 and 0.4)
    (2.5,3.5) ellipse (1 and 0.4);

\draw[fill=gray!50, dashed] 
    (1,0) arc [start angle=0, end angle=180, x radius=1, y radius=0.4]
    (3.5,0) arc [start angle=0, end angle=180, x radius=1, y radius=0.4];

\draw[fill=gray!50] 
    (1,0) arc [start angle=0, end angle=-180, x radius=1, y radius=0.4]
    (-1,0) arc [start angle=180, end angle=160, x radius=1, y radius=0.4]
    (3.5,0) arc [start angle=0, end angle=-180, x radius=1, y radius=0.4]
    (3.5,0) arc [start angle=0, end angle=20, x radius=1, y radius=0.4];

\draw
    (0.975,-0.09) -- (3.475,3.41)
    (-0.975,0.09) -- (1.525,3.59);

\begin{scope}
    \clip
        (0.975,-0.09) -- (3.475,3.41) -- (1.525,3.59) -- (-0.975,0.09);
    \draw[dashed]
        (3.475,0.09) -- (0.975,3.59)
        (1.525,-0.09) -- (-0.975,3.41);
\end{scope}

\node[scale=1] at (0,-0.8) {$|\psi\rangle$};
\node[scale=1] at (0,4.3) {$\langle\phi|$};
\node[scale=1] at (2.5,-0.8) {$|\phi\rangle$};
\node[scale=1] at (2.5,4.3) {$\langle\psi|$};

\end{tikzpicture}

%% file: tikz/macaroni.tex
\begin{tikzpicture}[very thick, scale=0.7]

\fill[fill=gray!50] 
    plot [smooth, tension=1.75] coordinates {(-1,0) (1.25,1.5) (3.5,0)} -- (3.5,0) arc [start angle=0, end angle=-180, x radius=1, y radius=0.4] -- plot [smooth, tension=1.75] coordinates {(1.5,0) (1.25,0.25) (1,0)} -- (1,0) arc [start angle=0, end angle=-180, x radius=1, y radius=0.4]
    plot [smooth, tension=1.75] coordinates {(-1,3.5) (1.25,2) (3.5,3.5)} -- (3.5,3.5) arc [start angle=0, end angle=-180, x radius=1, y radius=0.4] -- plot [smooth, tension=1.75] coordinates {(1.5,3.5) (1.25,3.25) (1,3.5)} -- (1,3.5) arc [start angle=0, end angle=-180, x radius=1, y radius=0.4];

\draw[fill=gray!85] 
    (0,3.5) ellipse (1 and 0.4)
    (2.5,3.5) ellipse (1 and 0.4);

\draw[fill=gray!50] 
    (1,0) arc [start angle=0, end angle=-180, x radius=1, y radius=0.4]
    (3.5,0) arc [start angle=0, end angle=-180, x radius=1, y radius=0.4];

\draw[fill=gray!50, dashed] 
    (1,0) arc [start angle=0, end angle=180, x radius=1, y radius=0.4]
    (3.5,0) arc [start angle=0, end angle=180, x radius=1, y radius=0.4];

\draw
    plot [smooth, tension=1.75] coordinates {(-1,0) (1.25,1.5) (3.5,0)}
    plot [smooth, tension=1.75] coordinates {(-1,3.5) (1.25,2) (3.5,3.5)}
    plot [smooth, tension=1.75] coordinates {(1,0) (1.25,0.25) (1.5,0)}
    plot [smooth, tension=1.75] coordinates {(1,3.5) (1.25,3.25) (1.5,3.5)};

\node[scale=1] at (0,-0.8) {$|\psi\rangle$};
\node[scale=1] at (0,4.3) {$\langle\phi|$};
\node[scale=1] at (2.5,-0.8) {$|\phi\rangle$};
\node[scale=1] at (2.5,4.3) {$\langle\psi|$};

\end{tikzpicture}

%% file: tikz/closed_ip_first.tex
\begin{tikzpicture}[very thick, scale=1]

\fill[gray!50] 
    (0,2) ellipse (0.65 and 0.26)
    (0,0) ellipse (1 and 0.4);

\fill[gray!50]
    plot [smooth,tension=1] coordinates {(0.65,2) (0.75,0.8) (0.977,0.085)} -- plot [smooth,tension=1] coordinates {(-0.977,0.085) (-0.75,0.8) (-0.65,2)} -- cycle;

\begin{scope}
    \clip (-1.1,0.1) rectangle (1.1,-0.5);
    \draw (0,0) ellipse (1 and 0.4);
\end{scope}
\draw[dashed] (0,0) ellipse (1 and 0.4);

\draw[line cap = round] 
    plot [smooth,tension=1] coordinates {(0.65,2) (0.75,0.8) (0.977,0.085)}
    plot [smooth,tension=1] coordinates {(-0.65,2) (-0.75,0.8) (-0.977,0.085)};

\draw[Kdarkgreen, ultra thick, fill=gray!85] (0,2) ellipse (0.65 and 0.26);

\draw[line width=4.25, Kblue, line cap = round] 
    (-0.65,2) arc [start angle=180, end angle=70, x radius=0.65, y radius = 0.26]
    (-0.65,2) arc [start angle=180, end angle=290, x radius=0.65, y radius = 0.26];

\node[scale=1.2, anchor=north] at (0,-0.4) {$|\psi\rangle$};

\node[scale=1.2, Kdarkgreen] at (0.8,2.1) {$h$};
\node[scale=1.2, Kdarkgreen] at (0.8,2.65) {$h$};

\begin{scope}[yscale=-1, shift={(0,-4.75)}]
    \fill[gray!50]
        plot [smooth,tension=1] coordinates {(0.65,2) (0.75,0.8) (0.977,0.085)} -- plot [smooth,tension=1] coordinates {(-0.977,0.085) (-0.75,0.8) (-0.65,2)} -- (0,2.1) -- cycle;
\end{scope}

\draw[fill=gray!85] (0,4.75) ellipse (1 and 0.4);

\begin{scope}[yscale=-1, shift={(0,-4.75)}]
\draw[line cap = round] 
    plot [smooth,tension=1] coordinates {(0.65,2) (0.75,0.8) (0.977,0.085)}
    plot [smooth,tension=1] coordinates {(-0.65,2) (-0.75,0.8) (-0.977,0.085)};
\end{scope}

\begin{scope}
    \clip (-1,2.4) rectangle (1,2.75);
    \draw[Kdarkgreen, ultra thick, fill=gray!50] (0,2.75) ellipse (0.65 and 0.26);
\end{scope}
\draw[Kdarkgreen, ultra thick, dashed] (0.65,2.75) arc [start angle=0, end angle=70, x radius=0.65, y radius = 0.26];

\node[scale=1.2, anchor=east, Kblue] at (-0.65, 2.75) {ob};

\draw[line width=4.25, Kblue, line cap = round, loosely dashed] 
    (-0.65,2.75) arc [start angle=180, end angle=70, x radius=0.65, y radius = 0.26];
\draw[line width=4.25, Kblue, line cap = round]
    (-0.65,2.75) arc [start angle=180, end angle=290, x radius=0.65, y radius = 0.26];

\node[scale=1.2,anchor=south] at (0,5.15) {$\langle\phi|$};

\end{tikzpicture}

%% file: tikz/closed_ip_second.tex
\begin{tikzpicture}[very thick, scale=1]

\fill[gray!50] 
    (0,2) ellipse (0.65 and 0.26)
    (0,0) ellipse (1 and 0.4);

\fill[gray!50]
    plot [smooth,tension=1] coordinates {(0.65,2) (0.75,0.8) (0.977,0.085)} -- plot [smooth,tension=1] coordinates {(-0.977,0.085) (-0.75,0.8) (-0.65,2)} -- cycle;

\begin{scope}
    \clip (-1.1,0.1) rectangle (1.1,-0.5);
    \draw (0,0) ellipse (1 and 0.4);
\end{scope}
\draw[dashed] (0,0) ellipse (1 and 0.4);

\draw[line cap = round] 
    plot [smooth,tension=1] coordinates {(0.65,2) (0.75,0.8) (0.977,0.085)}
    plot [smooth,tension=1] coordinates {(-0.65,2) (-0.75,0.8) (-0.977,0.085)};

\draw[Kdarkgreen, ultra thick, fill=gray!85] (0,2) ellipse (0.65 and 0.26);

\draw[line width=4.25, Kblue, line cap = round] 
    (-0.65,2) arc [start angle=180, end angle=70, x radius=0.65, y radius = 0.26]
    (-0.65,2) arc [start angle=180, end angle=290, x radius=0.65, y radius = 0.26];

\node[scale=1.2, anchor=north] at (0,-0.4) {$|\psi\rangle$};

\node[scale=1.2, Kdarkgreen] at (0.9,2.1) {$h_2$};
\node[scale=1.2, Kdarkgreen] at (0.9,2.65) {$h_1$};

\begin{scope}[yscale=-1, shift={(0,-4.75)}]
    \fill[gray!50]
        plot [smooth,tension=1] coordinates {(0.65,2) (0.75,0.8) (0.977,0.085)} -- plot [smooth,tension=1] coordinates {(-0.977,0.085) (-0.75,0.8) (-0.65,2)} -- (0,2.1) -- cycle;
\end{scope}

\draw[fill=gray!85] (0,4.75) ellipse (1 and 0.4);

\begin{scope}[yscale=-1, shift={(0,-4.75)}]
\draw[line cap = round] 
    plot [smooth,tension=1] coordinates {(0.65,2) (0.75,0.8) (0.977,0.085)}
    plot [smooth,tension=1] coordinates {(-0.65,2) (-0.75,0.8) (-0.977,0.085)};
\end{scope}

\begin{scope}
    \clip (-1,2.4) rectangle (1,2.75);
    \draw[Kdarkgreen, ultra thick, fill=gray!50] (0,2.75) ellipse (0.65 and 0.26);
\end{scope}
\draw[Kdarkgreen, ultra thick, dashed] (0.65,2.75) arc [start angle=0, end angle=70, x radius=0.65, y radius = 0.26];

\node[scale=1.2, anchor=east, Kblue] at (-0.65, 2.75) {ob};

\draw[line width=4.25, Kblue, line cap = round, loosely dashed] 
    (-0.65,2.75) arc [start angle=180, end angle=70, x radius=0.65, y radius = 0.26];
\draw[line width=4.25, Kblue, line cap = round]
    (-0.65,2.75) arc [start angle=180, end angle=290, x radius=0.65, y radius = 0.26];

\node[scale=1.2,anchor=south] at (0,5.15) {$\langle\phi|$};


\begin{scope}[shift={(2.5,0)}]
    \fill[gray!50] 
        (0,2) ellipse (0.65 and 0.26)
        (0,0) ellipse (1 and 0.4);
    
    \fill[gray!50]
        plot [smooth,tension=1] coordinates {(0.65,2) (0.75,0.8) (0.977,0.085)} -- plot [smooth,tension=1] coordinates {(-0.977,0.085) (-0.75,0.8) (-0.65,2)} -- cycle;
    
    \begin{scope}
        \clip (-1.1,0.1) rectangle (1.1,-0.5);
        \draw (0,0) ellipse (1 and 0.4);
    \end{scope}
    \draw[dashed] (0,0) ellipse (1 and 0.4);
    
    \draw[line cap = round] 
        plot [smooth,tension=1] coordinates {(0.65,2) (0.75,0.8) (0.977,0.085)}
        plot [smooth,tension=1] coordinates {(-0.65,2) (-0.75,0.8) (-0.977,0.085)};
    
    \draw[Kdarkgreen, ultra thick, fill=gray!85] (0,2) ellipse (0.65 and 0.26);
    
    \draw[line width=4.25, Kblue, line cap = round] 
        (-0.65,2) arc [start angle=180, end angle=70, x radius=0.65, y radius = 0.26]
        (-0.65,2) arc [start angle=180, end angle=290, x radius=0.65, y radius = 0.26];
    
    \node[scale=1.2, anchor=north] at (0,-0.4) {$|\phi\rangle$};
    
    \node[scale=1.2, Kdarkgreen] at (0.9,2.1) {$h_4$};
    \node[scale=1.2, Kdarkgreen] at (0.9,2.65) {$h_3$};
    
    \begin{scope}[yscale=-1, shift={(0,-4.75)}]
        \fill[gray!50]
            plot [smooth,tension=1] coordinates {(0.65,2) (0.75,0.8) (0.977,0.085)} -- plot [smooth,tension=1] coordinates {(-0.977,0.085) (-0.75,0.8) (-0.65,2)} -- (0,2.1) -- cycle;
    \end{scope}
    
    \draw[fill=gray!85] (0,4.75) ellipse (1 and 0.4);
    
    \begin{scope}[yscale=-1, shift={(0,-4.75)}]
    \draw[line cap = round] 
        plot [smooth,tension=1] coordinates {(0.65,2) (0.75,0.8) (0.977,0.085)}
        plot [smooth,tension=1] coordinates {(-0.65,2) (-0.75,0.8) (-0.977,0.085)};
    \end{scope}
    
    \begin{scope}
        \clip (-1,2.4) rectangle (1,2.75);
        \draw[Kdarkgreen, ultra thick, fill=gray!50] (0,2.75) ellipse (0.65 and 0.26);
    \end{scope}
    \draw[Kdarkgreen, ultra thick, dashed] (0.65,2.75) arc [start angle=0, end angle=70, x radius=0.65, y radius = 0.26];
    
    \draw[line width=4.25, Kblue, line cap = round, loosely dashed] 
        (-0.65,2.75) arc [start angle=180, end angle=70, x radius=0.65, y radius = 0.26];
    \draw[line width=4.25, Kblue, line cap = round]
        (-0.65,2.75) arc [start angle=180, end angle=290, x radius=0.65, y radius = 0.26];
    
    \node[scale=1.2,anchor=south] at (0,5.15) {$\langle\psi|$};
\end{scope}

\end{tikzpicture}

%% file: tikz/rigatoni_mod.tex
\begin{tikzpicture}[very thick, scale=1.2]

\fill[gray!50] 
    plot [smooth, tension=1] coordinates {(-1.25,1) (-1,0) (-1.25,-1)} --
    plot [smooth, tension=1] coordinates {(-1.25,-1) (-1,-1.05) (-0.5,-0.95) (0,-1.05) (0.5,-0.95) (1,-1.05) (1.25,-1)} --
    plot [smooth, tension=1] coordinates {(1.25,-1) (1,0) (1.25,1)} -- 
    plot [smooth, tension=1] coordinates {(1.25,1) (1,1.05) (0.5,0.95) (0,1.05) (-0.5,0.95) (-1,1.05) (-1.25,1)};

\draw[line cap = round] plot [smooth, tension=1] coordinates {(-1.25,1) (-1,0) (-1.25,-1)};

\draw[line cap = round] plot [smooth, tension=1] coordinates {(1.25,1) (1,0) (1.25,-1)};
\fill[white] (0.6,0.47) rectangle (1.1,-0.47);

\draw[ultra thick, Kblue, fill=gray!85] plot [smooth,tension=1] coordinates {({cos(70)},{0.4*sin(70)}) (1.06,-0.5) ({cos(70)},{-0.4*sin(70)})};

\draw[ultra thick, Kblue, fill=gray!50] plot [smooth,tension=1] coordinates {({cos(70)},{0.4*sin(70)}) (1.06,0.5) ({cos(70)},{-0.4*sin(70)})};

\begin{scope}
    \clip plot [smooth,tension=1] coordinates {({cos(70)},{0.4*sin(70)}) (1,0.5) ({cos(70)},{-0.4*sin(70)})} -- cycle;
    \draw[ultra thick, Kblue, dashed] plot [smooth,tension=1] coordinates {({cos(70)},{0.4*sin(70)}) (1,-0.5) ({cos(70)},{-0.4*sin(70)})};
\end{scope}

\draw[line width=4.25, Kblue, line cap = round, loosely dashed] 
    (-1,0) arc [start angle=180, end angle=70, x radius=1, y radius = 0.4];
\draw[line width=4.25, Kblue, line cap = round]
    (-1,0) arc [start angle=180, end angle=290, x radius=1, y radius = 0.4];

\node[scale=1.2, anchor=east, Kblue] at (-1, 0) {ob};

\draw[Kblue, <->] plot [smooth] coordinates {(1,-0.3) (1.1,0) (1,0.3)};


\begin{scope}[xscale=-1, shift={(-3,0)}]
    \fill[gray!50] 
        plot [smooth, tension=1] coordinates {(-1.25,1) (-1,0) (-1.25,-1)} --
        plot [smooth, tension=1] coordinates {(-1.25,-1) (-1,-1.05) (-0.5,-0.95) (0,-1.05) (0.5,-0.95) (1,-1.05) (1.25,-1)} --
        plot [smooth, tension=1] coordinates {(1.25,-1) (1,0) (1.25,1)} -- 
        plot [smooth, tension=1] coordinates {(1.25,1) (1,1.05) (0.5,0.95) (0,1.05) (-0.5,0.95) (-1,1.05) (-1.25,1)};
    
    \draw[line cap = round] plot [smooth, tension=1] coordinates {(-1.25,1) (-1,0) (-1.25,-1)};
    
    \draw[line cap = round] plot [smooth, tension=1] coordinates {(1.25,1) (1,0) (1.25,-1)};
    \fill[white] (0.6,0.47) rectangle (1.1,-0.47);
    
    \draw[ultra thick, Kblue, fill=gray!85] plot [smooth,tension=1] coordinates {({cos(70)},{0.4*sin(70)}) (1.06,-0.5) ({cos(70)},{-0.4*sin(70)})};
    
    \draw[ultra thick, Kblue, fill=gray!50] plot [smooth,tension=1] coordinates {({cos(70)},{0.4*sin(70)}) (1.06,0.5) ({cos(70)},{-0.4*sin(70)})};
    
    \begin{scope}
        \clip plot [smooth,tension=1] coordinates {({cos(70)},{0.4*sin(70)}) (1,0.5) ({cos(70)},{-0.4*sin(70)})} -- cycle;
        \draw[ultra thick, Kblue, dashed] plot [smooth,tension=1] coordinates {({cos(70)},{0.4*sin(70)}) (1,-0.5) ({cos(70)},{-0.4*sin(70)})};
    \end{scope}
    
    \draw[line width=4.25, Kblue, line cap = round, loosely dashed] 
        (-1,0) arc [start angle=180, end angle=70, x radius=1, y radius = 0.4];
    \draw[line width=4.25, Kblue, line cap = round]
        (-1,0) arc [start angle=180, end angle=290, x radius=1, y radius = 0.4];
    
    \draw[Kblue, <->] plot [smooth] coordinates {(1,-0.3) (1.1,0) (1,0.3)};
\end{scope}

\end{tikzpicture}

%% file: tikz/penne_mod.tex
\begin{tikzpicture}[very thick, scale=1.2]

\fill[gray!50] 
    plot [smooth, tension=1] coordinates {(-1.25,1) (-1,0) (-1.25,-1)} --
    plot [smooth, tension=1] coordinates {(-1.25,-1) (-1,-1.05) (-0.5,-0.95) (0,-1.05) (0.5,-0.95) (1,-1.05) (1.25,-1)} --
    plot [smooth, tension=1] coordinates {(1.25,-1) (1,0) (1.25,1)} -- 
    plot [smooth, tension=1] coordinates {(1.25,1) (1,1.05) (0.5,0.95) (0,1.05) (-0.5,0.95) (-1,1.05) (-1.25,1)};

\draw[line cap = round] plot [smooth, tension=1] coordinates {(-1.25,1) (-1,0) (-1.25,-1)};

\draw[line cap = round] plot [smooth, tension=1] coordinates {(1.25,1) (1,0) (1.25,-1)};
\fill[white] (0.6,0.47) rectangle (1.1,-0.47);

\draw[ultra thick, Kpurple, fill=gray!85] plot [smooth,tension=1] coordinates {({cos(70)},{0.4*sin(70)}) (1.06,-0.5) ({cos(70)},{-0.4*sin(70)})};

\draw[ultra thick, Kpurple, fill=gray!50] plot [smooth,tension=1] coordinates {({cos(70)},{0.4*sin(70)}) (1.06,0.5) ({cos(70)},{-0.4*sin(70)})};

\begin{scope}
    \clip plot [smooth,tension=1] coordinates {({cos(70)},{0.4*sin(70)}) (1,0.5) ({cos(70)},{-0.4*sin(70)})} -- cycle;
    \draw[ultra thick, Kpurple, dashed] plot [smooth,tension=1] coordinates {({cos(70)},{0.4*sin(70)}) (1,-0.5) ({cos(70)},{-0.4*sin(70)})};
\end{scope}

\draw[line width=4.25, Kblue, line cap = round, loosely dashed] 
    (-1,0) arc [start angle=180, end angle=70, x radius=1, y radius = 0.4];
\draw[line width=4.25, Kblue, line cap = round]
    (-1,0) arc [start angle=180, end angle=290, x radius=1, y radius = 0.4];

\node[scale=1.2, anchor=east, Kblue] at (-1,0) {ob};


\begin{scope}[xscale=-1, shift={(-3,0)}]
    \fill[gray!50] 
        plot [smooth, tension=1] coordinates {(-1.25,1) (-1,0) (-1.25,-1)} --
        plot [smooth, tension=1] coordinates {(-1.25,-1) (-1,-1.05) (-0.5,-0.95) (0,-1.05) (0.5,-0.95) (1,-1.05) (1.25,-1)} --
        plot [smooth, tension=1] coordinates {(1.25,-1) (1,0) (1.25,1)} -- 
        plot [smooth, tension=1] coordinates {(1.25,1) (1,1.05) (0.5,0.95) (0,1.05) (-0.5,0.95) (-1,1.05) (-1.25,1)};
    
    \draw[line cap = round] plot [smooth, tension=1] coordinates {(-1.25,1) (-1,0) (-1.25,-1)};
    
    \draw[line cap = round] plot [smooth, tension=1] coordinates {(1.25,1) (1,0) (1.25,-1)};
    \fill[white] (0.6,0.47) rectangle (1.1,-0.47);
    
    \draw[ultra thick, Kpurple, fill=gray!85] plot [smooth,tension=1] coordinates {({cos(70)},{0.4*sin(70)}) (1.06,-0.5) ({cos(70)},{-0.4*sin(70)})};
    
    \draw[ultra thick, Kpurple, fill=gray!50] plot [smooth,tension=1] coordinates {({cos(70)},{0.4*sin(70)}) (1.06,0.5) ({cos(70)},{-0.4*sin(70)})};
    
    \begin{scope}
        \clip plot [smooth,tension=1] coordinates {({cos(70)},{0.4*sin(70)}) (1,0.5) ({cos(70)},{-0.4*sin(70)})} -- cycle;
        \draw[ultra thick, Kpurple, dashed] plot [smooth,tension=1] coordinates {({cos(70)},{0.4*sin(70)}) (1,-0.5) ({cos(70)},{-0.4*sin(70)})};
    \end{scope}
    
    \draw[line width=4.25, Kblue, line cap = round, loosely dashed] 
        (-1,0) arc [start angle=180, end angle=70, x radius=1, y radius = 0.4];
    \draw[line width=4.25, Kblue, line cap = round]
        (-1,0) arc [start angle=180, end angle=290, x radius=1, y radius = 0.4];
\end{scope}

\draw[Kpurple, <->] plot [smooth] coordinates {(1,-0.3) (2,0.3)};
\draw[Kpurple, <->] plot [smooth] coordinates {(1,0.3) (2,-0.3)};

\end{tikzpicture}

%% file: tikz/macaroni_mod.tex
\begin{tikzpicture}[very thick, scale=1.2]

\fill[gray!50] 
    plot [smooth, tension=1] coordinates {(-1.25,1) (-1,0) (-1.25,-1)} --
    plot [smooth, tension=1] coordinates {(-1.25,-1) (-1,-1.05) (-0.5,-0.95) (0,-1.05) (0.5,-0.95) (1,-1.05) (1.25,-1)} --
    plot [smooth, tension=1] coordinates {(1.25,-1) (1,0) (1.25,1)} -- 
    plot [smooth, tension=1] coordinates {(1.25,1) (1,1.05) (0.5,0.95) (0,1.05) (-0.5,0.95) (-1,1.05) (-1.25,1)};

\draw[line cap = round] plot [smooth, tension=1] coordinates {(-1.25,1) (-1,0) (-1.25,-1)};

\draw[line cap = round] plot [smooth, tension=1] coordinates {(1.25,1) (1,0) (1.25,-1)};
\fill[white] (0.6,0.47) rectangle (1.1,-0.47);

\draw[ultra thick, Korange, fill=gray!85] plot [smooth,tension=1] coordinates {({cos(70)},{0.4*sin(70)}) (1.06,-0.5) ({cos(70)},{-0.4*sin(70)})};

\draw[ultra thick, Korange, fill=gray!50] plot [smooth,tension=1] coordinates {({cos(70)},{0.4*sin(70)}) (1.06,0.5) ({cos(70)},{-0.4*sin(70)})};

\begin{scope}
    \clip plot [smooth,tension=1] coordinates {({cos(70)},{0.4*sin(70)}) (1,0.5) ({cos(70)},{-0.4*sin(70)})} -- cycle;
    \draw[ultra thick, Korange, dashed] plot [smooth,tension=1] coordinates {({cos(70)},{0.4*sin(70)}) (1,-0.5) ({cos(70)},{-0.4*sin(70)})};
\end{scope}

\draw[line width=4.25, Kblue, line cap = round, loosely dashed] 
    (-1,0) arc [start angle=180, end angle=70, x radius=1, y radius = 0.4];
\draw[line width=4.25, Kblue, line cap = round]
    (-1,0) arc [start angle=180, end angle=290, x radius=1, y radius = 0.4];

\node[scale=1.2, anchor=east, Kblue] at (-1,0) {ob};


\begin{scope}[xscale=-1, shift={(-3,0)}]
    \fill[gray!50] 
        plot [smooth, tension=1] coordinates {(-1.25,1) (-1,0) (-1.25,-1)} --
        plot [smooth, tension=1] coordinates {(-1.25,-1) (-1,-1.05) (-0.5,-0.95) (0,-1.05) (0.5,-0.95) (1,-1.05) (1.25,-1)} --
        plot [smooth, tension=1] coordinates {(1.25,-1) (1,0) (1.25,1)} -- 
        plot [smooth, tension=1] coordinates {(1.25,1) (1,1.05) (0.5,0.95) (0,1.05) (-0.5,0.95) (-1,1.05) (-1.25,1)};
    
    \draw[line cap = round] plot [smooth, tension=1] coordinates {(-1.25,1) (-1,0) (-1.25,-1)};
    
    \draw[line cap = round] plot [smooth, tension=1] coordinates {(1.25,1) (1,0) (1.25,-1)};
    \fill[white] (0.6,0.47) rectangle (1.1,-0.47);
    
    \draw[ultra thick, Korange, fill=gray!85] plot [smooth,tension=1] coordinates {({cos(70)},{0.4*sin(70)}) (1.06,-0.5) ({cos(70)},{-0.4*sin(70)})};
    
    \draw[ultra thick, Korange, fill=gray!50] plot [smooth,tension=1] coordinates {({cos(70)},{0.4*sin(70)}) (1.06,0.5) ({cos(70)},{-0.4*sin(70)})};
    
    \begin{scope}
        \clip plot [smooth,tension=1] coordinates {({cos(70)},{0.4*sin(70)}) (1,0.5) ({cos(70)},{-0.4*sin(70)})} -- cycle;
        \draw[ultra thick, Korange, dashed] plot [smooth,tension=1] coordinates {({cos(70)},{0.4*sin(70)}) (1,-0.5) ({cos(70)},{-0.4*sin(70)})};
    \end{scope}
    
    \draw[line width=4.25, Kblue, line cap = round, loosely dashed] 
        (-1,0) arc [start angle=180, end angle=70, x radius=1, y radius = 0.4];
    \draw[line width=4.25, Kblue, line cap = round]
        (-1,0) arc [start angle=180, end angle=290, x radius=1, y radius = 0.4];
\end{scope}

\draw[Korange, <->] plot [smooth] coordinates {(1,0.3) (1.5,0.1) (2,0.3)};
\draw[Korange, <->] plot [smooth] coordinates {(1,-0.3) (1.5,-0.1) (2,-0.3)};

\end{tikzpicture}

%% file: tikz/pre-average.tex
\begin{tikzpicture}[thick,scale=0.8]

\node[scale=1.3, anchor=east] at (-1,1.75) {$d_{b_1}^2 d_{b_2}^2 d_{e_1} d_{e_2} \displaystyle\int dO_1$};

\draw[fill=gray!20] (0,0) rectangle (2,1);
\node[scale=1.5] at (1,0.5) {$O_1$};
\draw[fill=gray!20] (2.5,0) rectangle (4.5,1);
\node[scale=1.5] at (3.5,0.5) {$O_2$};

\draw (1,-0.25) -- (1,0);
\node[scale=1, anchor=north] at (1,-0.25) {$|0\rangle$};
\draw (3.5,-0.25) -- (3.5,0);
\node[scale=1, anchor=north] at (3.5,-0.25) {$|0\rangle$};

\draw[Kred] (-0.5,-0.5) -- (-0.5,1.25) -- (0,1.5) -- (0.5,1.25) -- (0.5,1);
\node[scale=1.2, anchor=north] at (-0.5,-0.5) {$|\psi\rangle_{b_1}$};

\draw[Kred] (1.5,1) -- (1.5,1.25) -- (2.25,1.5) -- (3,1.25) -- (3,1);

\draw[Kred] (4,1) -- (4,1.25) -- (4.5,1.5) -- (5,1.25) -- (5,-0.5);
\node[scale=1.2, anchor=north] at (5,-0.5) {$|\psi'\rangle_{b_2}$};

\begin{scope}[yscale=-1,shift={(0,-3.5)}]

    \draw[fill=gray!20] (0,0) rectangle (2,1);
    \node[scale=1.5] at (1,0.5) {$O_1^\intercal$};
    \draw[fill=gray!20] (2.5,0) rectangle (4.5,1);
    \node[scale=1.5] at (3.5,0.5) {$O_2^\intercal$};
    
    \draw (1,-0.25) -- (1,0);
    \node[scale=1, anchor=south] at (1,-0.25) {$\langle0|$};
    \draw (3.5,-0.25) -- (3.5,0);
    \node[scale=1, anchor=south] at (3.5,-0.25) {$\langle0|$};
    
    \draw[Kred] (-0.5,-0.5) -- (-0.5,1.25) -- (0,1.5) -- (0.5,1.25) -- (0.5,1);
    \node[scale=1.2, anchor=south] at (-0.5,-0.5) {$\langle\psi|_{b_1}$};
    
    \draw[Kred] (1.5,1) -- (1.5,1.25) -- (2.25,1.5) -- (3,1.25) -- (3,1);
    
    \draw[Kred] (4,1) -- (4,1.25) -- (4.5,1.5) -- (5,1.25) -- (5,-0.5);
    \node[scale=1.2, anchor=south] at (5,-0.5) {$\langle\psi'|_{b_2}$};

\end{scope}

\begin{scope}[shift={(9,0)}]

    \node[scale=1.3] at (-2.75,1.75) {$=$};
    \node[scale=1.3, anchor=east] at (-1,1.75) {$d_{b_2}^2$};
    
    \draw[fill=gray!20] (0,0) rectangle (2,1);
    \node[scale=1.5] at (1,0.5) {$O_2$};
    
    \draw (1,-0.25) -- (1,0);
    \node[scale=1, anchor=north] at (1,-0.25) {$|0\rangle$};
    
    \draw[Kred] (-0.75,-0.5) -- (-0.75,2);
    \node[scale=1.2, anchor=north] at (-0.75,-0.5) {$|\psi\rangle_{b_1}$};
    
    \draw[Kred] (0.5,1) -- (0.5,2);
    
    \draw[Kred] (1.5,1) -- (1.5,1.25) -- (2,1.5) -- (2.5,1.25) -- (2.5,-0.5);
    \node[scale=1.2, anchor=north] at (2.5,-0.5) {$|\psi'\rangle_{b_2}$};

\end{scope}

\begin{scope}[yscale=-1,shift={(9,-3.5)}]
    
    \draw[fill=gray!20] (0,0) rectangle (2,1);
    \node[scale=1.5] at (1,0.5) {$O_2^\intercal$};
    
    \draw (1,-0.25) -- (1,0);
    \node[scale=1, anchor=south] at (1,-0.25) {$\langle0|$};
    
    \draw[Kred] (-0.75,-0.5) -- (-0.75,1.5);
    \node[scale=1.2, anchor=south] at (-0.75,-0.5) {$\langle\psi|_{b_1}$};
    
    \draw[Kred] (0.5,1) -- (0.5,1.5);
    
    \draw[Kred] (1.5,1) -- (1.5,1.25) -- (2,1.5) -- (2.5,1.25) -- (2.5,-0.5);
    \node[scale=1.2, anchor=south] at (2.5,-0.5) {$\langle\psi'|_{b_2}$};

\end{scope}

\end{tikzpicture}